\documentclass[12pt,preprint]{aastex}
\usepackage{emulateapj5}
\begin{document}

\newcommand{\MITRGB}{\rm M_{I}(TRGB)}
\newcommand{\MI}{\rm M_{I}}

\shorttitle{Star Formation History and the TRGB}
\shortauthors{Barker, Sarajedini, \& Harris}

\slugcomment{Accepted for publication in The Astrophysical Journal
Supplements}

\title{Variations in Star Formation History and the Red Giant Branch Tip}

\author{Michael K. Barker\altaffilmark{1}}
\author{Ata Sarajedini\altaffilmark{1}}
\and
\author{Jason Harris\altaffilmark{2,3}}

\altaffiltext{1}{Department of Astronomy, University of Florida,
  Gainesville, FL 32611; mbarker@astro.ufl.edu; ata@astro.ufl.edu}
%\email{mbarker@astro.ufl.edu}
%\email{ata@astro.ufl.edu}

\altaffiltext{2}{Steward Observatory, 933 N. Cherry Ave., Tucson, AZ 85721;
jharris@as.arizona.edu}
%\email{jharris@as.arizona.edu}

\altaffiltext{3}{Hubble Fellow}

\begin{abstract} 

We examine the reliability of the tip of the red giant branch (TRGB)
as a distance indicator for stellar populations with 
different star formation histories (SFHs) when 
photometric errors and completeness corrections at the TRGB
are small.  In general, the 
TRGB-distance method is insensitive to the shape of the 
SFH except when it produces a stellar 
population with a significant component undergoing the red giant 
branch phase transition.
The I-band absolute magnitude of the TRGB for the middle and
late stages of this transition ($\sim 1.3-1.7$ Gyr) is 
several tenths of a magnitude fainter than the canonical value 
of $\rm M_I \approx -4.0$.  
If more than $\sim 30\%$ of all stars formed over the lifetime
of the Universe are formed at these
ages, then the distance could be overestimated by
$\sim 10-25\%$.
Similarly, the TRGB-distance method is insensitive to the 
metallicity distribution of stars formed except when the
average metallicity is greater than $\rm \langle[Fe/H]\rangle = -0.3$.
If more than $\sim 70\%$ of all stars formed have [Fe/H] $> -0.3$, 
the distance could be overestimated by $\sim 10-45\%$.
We find that two observable quantities, the height of the 
discontinuity in the luminosity function at the TRGB and the
median $\rm (V-I)_0$ at $\rm M_I = -3.5$ can be used to test
if the aforementioned age and metallicity conditions are met.

\end{abstract}

\keywords{galaxies: dwarf --- galaxies: stellar content --- 
Local Group --- galaxies: distances and redshifts}

\section{Introduction}

The tip of the red giant branch (TRGB) marks the onset of helium fusion in
stars.  For old, low-mass stars this is an 
explosive event called the helium flash.  During the 
helium flash these stars undergo a quick readjustment
of their interiors before they begin helium burning on the horizontal branch.
Therefore, the TRGB is manifested as a sharp discontinuity in the RGB 
luminosity function (LF).  Theoretical models and observational data 
agree that the I-band absolute magnitude of the TRGB is 
located at $\MITRGB \approx -4.0\ \pm\
0.1$ for metallicities in the range $-2.2 \lesssim {\rm[Fe/H]} \lesssim 
-0.7$ and ages between 2 and 15 Gyr (Da Costa \& Armandroff 1990; 
Lee, Freedman, \& Madore 1993; Bellazzini, Ferraro, \& Pancino 2001).  
The reason for this is twofold.  First, the mass of the degenerate 
helium core and hence, the stellar luminosity, is
approximately constant in this age range.  
Second, the stellar spectrum peaks near the I-band in these age and 
metallicity ranges making $\MITRGB$ approximately
constant.  Therefore, $\MITRGB$ can be used as a 
standard candle for measuring distances to resolved stellar
populations.

Numerous studies have taken advantage of the TRGB standard candle
to measure distances to galaxies in the Local Group and beyond
(e.g. M$\rm \acute{e}$ndez et al.\ 2002;
Karachentsev et al.\ 2002; Sakai et al.\ 1997; Cioni et al.\ 2000; 
Sarajedini et al.\ 2002; Jerjen \& Rejkuba 2001; 
Ma$\rm \acute{i}$z-Apell$\rm \acute{a}$niz, Cieza, \& MacKenty 2002; 
Sakai, Zaritsky, \& Kennicutt 2000; Sakai, Madore, \& Freedman 1999).
Such distance measurments are important for calibrating
other distance measurement techniques, for measuring the
local velocity field (M$\rm \acute{e}$ndez et al. 2002; 
Karachentsev et al. 2002), and for determining accurate SFHs
of the galaxy stellar populations.  However, many of these
studies make an assumption that the target RGB populations
occupy the required age and metallicity ranges.

In this paper, we examine the effects of star formation
history (SFH) variations on the TRGB-distance method.  In particular,
we generate synthetic color-magnitude diagrams (CMDs) for stellar
populations with a range of ages and metallicities and
employ standard observational techniques to determine the
I magnitude of the TRGB.  Since the distance to each synthetic
population is known, we can quantitatively evaluate the 
systematic errors associated with the SFH variations on the
TRGB-distance method.  We define old stars as those having
ages $> 10$ Gyr, intermediate-age stars as those with ages 
between 2 and 7 Gyr, and young stars as those with ages 
$< 2$ Gyr.

\section{Methods}

\subsection{Synthetic color-magnitude diagrams}

To generate synthetic CMDs, 
we use the StarFISH software introduced by Harris \& Zaritsky (2001).  
Given an SFH, distance modulus, extinction law, reddening, initial mass
function (IMF) slope and binary fraction, this code uses theoretical
isochrones to generate photometry for artificial stellar
populations.  We adopt the Padova set of isochrones 
(Girardi et al.\ 2000) which cover metal abundances of Z = 0.0004, 0.001,
0.004, 0.008, 0.019, 0.030 and an age range of log(age/yr) = $7.80-10.25$.  
These isochrones include evolution from the main sequence
to either the thermally pulsing asymptotic giant branch (AGB) stage 
or carbon ignition so all of our synthetic CMDs contain AGB stars 
at luminosities higher than or comparable to that of the TRGB.
To increase metallicity resolution we linearly interpolate between each 
metallicity pair resulting in another set of isochrones with the 
same ages but 
metallicities of Z = 0.0007, 0.0025, 0.006, 0.0135, and 0.0245.  
For simplicity and clarity, we use a distance modulus of zero, Salpeter IMF,
binary fraction of 0.25, and zero reddening.  

The StarFISH software populates isochrones with stars in a
realistic manner; the probability of finding a star at any 
point along the isochrone depends on the IMF and on the
evolutionary timescale of that particular point.  To reproduce
observational effects, the stars added to an isochrone are 
probabilistically removed and scattered according to 
photometry-dependent completeness rate and error tables.

\subsection{Photometric errors and completeness}

Generating accurate synthetic CMDs requires a proper analysis of how 
photometric errors and completeness fraction change with magnitude and color.  
When studying a real stellar population, this analysis is done with 
artificial star tests in which stars of
known magnitudes and colors are placed in the original images and the data
reduction is repeated.  The completeness fraction is estimated from the 
number of recovered artificial stars whose colors and magnitudes 
are then compared to the input colors and magnitudes to estimate photometric
errors (e.g.\ Aparicio \& Gallart 1995).  

Because the TRGB is a relatively confined region of the CMD the 
photometric errors and completeness fraction do not change 
significantly over the region of interest.  Moreover, photometric 
errors are 
typically small in this region because it is usually one of the brightest 
features of a CMD.  Crowding also is generally not
severe in nearby dwarf galaxies and in the halos of LG galaxies where the 
TRGB-distance method is typically applied.  However, 
when applying the TRGB-distance method to galaxies beyond the LG
or in very crowded stellar regions nearby, the errors and completeness
rate can be much more important.  
Investigating the effect of SFH variations
on the TRGB in such cases is beyond the scope of the
present study.

Given the previous considerations, we choose to use simple analytic 
functions for the photometric errors
and completeness rate which are fairly typical of data taken with HST or 
ground-based telescopes (e.g.\ Aparicio \& Gallart 1995; Sarajedini 
et al.\ 2002).  These functions are plotted in Fig.\ 1
assuming the parameter values given below.
Photometric errors are modeled with an exponential function of the form,
\begin{equation}
\sigma(M) = \kappa e^{\tau M},
\end{equation}
where M is $\rm M_I$ or $\rm M_V$ and $\kappa$ and $\tau$ are 
constants.  Given the 
boundary conditions $\rm \sigma(M_I = -8.0) = 0.005$ and 
$\rm \sigma(M_I = 1.5) = 0.2$, these constants take the values
$\kappa = 0.11$ and $\tau = 0.39$.  
Completeness is modeled with the function,
\begin{equation}
f(M_V) = -\frac{2}{\pi} \arctan[\alpha(M_V-M_{V_0})],
\end{equation}
where $\alpha$ is a shape parameter that affects the steepness of the fall-off
and $\rm M_{V_0}$ is the magnitude at which the 
completeness equals 0\%.  For all our CMDs we use $\alpha$ = 2.0 and 
$\rm M_{V_0}$ = 1.5 which yield a completeness rate of 
50\% at $\rm M_V$ = 1.0.  

Each star is probabilistically removed according
to the completeness fraction at the star's $\rm M_V$ magnitude.  
If a star is kept,
its photometry is scattered by randomly drawing from a Gaussian distribution
with standard deviation equal to its error given by Equ.\ 1.

\subsection{Calibration of $\MITRGB$}

To use the TRGB as a distance indicator it must be calibrated by using
nearby stellar populations whose distances are known from other independent
techniques.  Several studies have done just that.  Da Costa \& Armandroff 
(1990) used the RGBs of eight
Galactic globular clusters to derive the I-band bolometric correction as a 
function of $\rm (V-I)_0$ color, the bolometric magnitude of the TRGB as a
function of [Fe/H], and [Fe/H] as a function of $\rm (V-I)_0$ color at 
$\rm M_I = -3.0$.  
Using a slightly revised version of the latter relation more appropriate for
galaxies, Lee et al.\ (1993) demonstrated that the TRGB method had the 
precision of primary distance indicators like Cepheids and RR Lyraes.  Most
recently, Bellazzini et al.\ (2001) derived a new empirical relation between
$\rm M_{I}(TRGB)$ and [Fe/H] based on the largest IR database of Galactic 
globular cluster RGBs and calibrated over a larger range of metallicities.

Since we are studying CMDs generated from 
theoretical isochrones, it is logical to use a theoretical rather than
empirical calibration of the TRGB.
In Fig.\ 2, we plot the I-band absolute magnitude of the TRGB as a function
of age for different metallicities.  These data are taken from the summary 
tables of the Padova isochrones.
For ages between 2 and 15 Gyr and $\rm -1.7 \leq [Fe/H] \leq -0.7$, 
$\MITRGB$ lies in 
the range $-4.0 \pm 0.05$, where 0.05 is the standard deviation.  
We use this theoretical estimate of $\MITRGB$ as 
a fiducial value when analyzing our synthetic CMDs.  

It is worthwhile to take a closer look at Fig.\ 2.  The 
qualitative behavior of $\MITRGB$ may be completely
explained by considering the results of Sweigart, Greggio, \& Renzini (1990).  
They examined in detail the dependence of 
the core mass on the total stellar mass at the TRGB.  They found
that for stars massive enough to avoid the degenerate conditions
that lead to the He flash, 
the core mass is approximately linearly proportional to the
total mass.  For low-mass stars, the core mass is completely 
degenerate and therefore insensitive to total mass.  But for stars in between
these two regimes, known as the RGB phase transition, 
the core mass increases quickly with decreasing total mass.  Because the
luminosity at the TRGB is so closely coupled to the core mass, it follows
the same pattern.  Since the total mass at the TRGB decreases with 
age, the luminosity has a similar dependence on age as it does on total
mass.
Finally, since $\rm M_I$ is proportional to luminosity, except when the 
metal abundance
is high enough that line-blanketing becomes significant in the I-band, 
$\rm M_I$ 
should depend on age the same way luminosity and core mass do.  

This is exactly what is
seen in Fig.\ 2.  Up to an age of $\sim 1$ Gyr, $\MITRGB$ 
becomes fainter with increasing age 
because the He
core reaches ignition before degeneracy sets in.  The RGB phase transition 
occurs between about 1 and 2 Gyr during which time the He core is partially
degenerate and $\MITRGB$ brightens very quickly with age.  
After 2 Gyr, the core is
completely degenerate and the basic stellar properties at the TRGB
like radius, mass, and
temperature change little with age.  For $\rm [Fe/H] > -0.7$, line
blanketing supresses $\MITRGB$.  

The exact age and duration of the RGB phase transition depends
on the details of the stellar models.  To demonstrate this 
point, in Fig.\ 3 we
compare the RGB phase transition of the 
Padova isochrone set (solid lines) with that of the $\rm Y^2$ set 
(dotted; Yi et al.\ 2001).  
The metallicities in panels (a), (b), (c), and (d)
are [Fe/H] = -1.7, -1.3, -0.7, and 0.0, respectively.  
We also show in panel (d) the Padova solar metallicity
isochrone with no convective overshoot (dashed).  
In general, there is 
good agreement between the $\rm Y^2$ and Padova isochrones 
except for the no overshoot curve in panel (d) which predicts
an earlier age and shorter duration for the transition.
This is in agreement 
with Ferraro et al.\ (1995) who found that models without
overshoot predict the transition to occur at $\sim 0.6 \pm 0.2$ Gyr
and last for $\sim 0.3$ Gyr.

To illustrate how the RGB phase transition could affect the 
TRGB-distance method, we show in Fig.\ 4 the evolution of the TRGB in the 
$\rm M_I-(V-I)_0$ plane from 0.1 Gyr to 
17.8 Gyr.  In this figure, ages of 1, 2, and 14 Gyr are marked
by triangles, asterisks, and squares, respectively.
For ages $\lesssim$ 0.14 Gyr, the TRGB is brighter than $\rm M_I = -4.0$ 
and it evolves to
fainter magnitudes until $\sim$ 1 Gyr when the core mass at the TRGB 
becomes partially degenerate.
At these young ages, the TRGB is substantially bluer than the 
TRGB of ages $>$ 2 Gyr.  Hence, it would simply add
background noise to a TRGB estimate which could be minimized by excluding 
stars bluer than a certain color, say $\rm (V-I)_0 \sim 1.1$ for
[Fe/H] = $-1.7$ and $-1.3$ or $\rm (V-I)_0 \sim 1.3$ for [Fe/H] = $-0.7$.
The RGB phase transition occurs approximately between 1 and 2 Gyr.  
If a stellar population is observed at some point during this stage, the 
TRGB will be fainter than $\rm M_I = -4.0$.  One could, in principle, measure 
the 
height of the RGB between the red clump (RC) and the TRGB to test if this was 
the case.  However, the RC is not always detected and
observational errors could prevent the use of the RGB length as an indicator
especially for the late stages of the RGB transition.  

To further illustrate our point,
in Fig.\ 5 we plot the RGBs for [Fe/H] = $-1.3$ and 
2, 1.78, 1.58, and 1.41 Gyr.  
The largest color difference between the RGBs 
at $\rm M_I = -3.0$ is $\sim 0.04$
and the color error at this magnitude is 
$\sigma_{\rm V-I} \sim 0.07$.
Therefore, under typical error conditions, 
these RGBs would be indistinguishable in the CMD of a galaxy
with an unkown distance modulus.
We will focus our attention on the ages of 1.58 and 1.78 Gyr 
when we construct our synthetic CMDs in later sections.

\subsection{Determination of $\MITRGB$}

For each synthetic CMD, we will measure $\MITRGB$ and compare it to
the calibration from the previous section.
At least six different techniques have been used in the literature 
to measure the magnitude of the TRGB.  Each method searches for
the discontinuity in the LF of RGB stars corresponding to the TRGB.
Lee et al.\ (1993) introduced the use of the zero-sum 
Sobel kernel, [$-2$, 0, +2],
whose convolution with the LF has a maximum at the largest discontinuity in
the LF.  Prior to their work, people estimated the TRGB by eye, making it 
difficult to reproduce and assign errors to their results.  A modified
Sobel filter was used by Sakai, Madore, \& Freedman (1996) with the definition
\begin{equation}
E(m) = \Phi(m+\bar{\sigma}) - \Phi(m-\bar{\sigma}),
\end{equation}
where $\bar{\sigma}$ is the average photometric error in the range m $\pm$ 
0.5 mag and $\Phi$(m) is a Gaussian-smoothed LF calculated as
\begin{equation}
\Phi(m) = \sum^{N}_{i=1} \frac{1}{\sqrt{2\pi}\sigma_i}\exp\left[-\frac{(m_i - m)^{2}}{2\sigma^{2}_{i}}\right].
\end{equation}
In this context, each star has a Gaussian probability distribution centered at
the star's magnitude, $m_i$, with dispersion, $\sigma_i$.  Stars
with small errors have more strongly peaked Gaussians than stars with large
errors.  The total LF is then a sum of all the stars' Gaussians.  

Motivated by
the sensitivity of the zero-sum Sobel filter to low-number counts, Frayn \& 
Gilmore (2003) introduced a third method to find the TRGB.  They fit a 
Gaussian cutoff to the LF in the region $\pm$1.0 mag of the Sobel filter's
determination of the TRGB.  Their method was found to give more reliable 
results in the low-number count regime.  Yet another method was used by 
Cioni et al.\ (2000) who estimated the TRGB to be the maximum of the second 
derivative of the binned LF plus a small systematic-error correction based on
a model for the intrinsic LF.  In addition to the Gaussian-smoothed Sobel
filter described above, M$\rm \acute{e}$ndez et al.\ (2002) estimated the TRGB
magnitude using the method of maximum likelihood.  They maximized 
the likelihood
function on a grid of three parameters, including the magnitude of the TRGB, 
that described the intrinsic LF of their data.
Finally, Sarajedini et al.\ (2002) used a simple slope-finding 
algorithm to estimate the TRGB magnitude.

All of the methods described above require {\it a priori} knowledge of 
the general location of the TRGB
because there are other features of the LF, such as the
red clump, AGB tip, AGB bump (Bellazzini 2002), 
and RGB bump (Bellazzini 2002; 
Monaco et al.\ 2002; Fusi Pecci 1990) 
which exhibit discontinuities that can be comparable in size to the 
TRGB.  Each method's accuracy and precision also depend on 
the signal-to-noise
ratio at the TRGB, field-star contamination, and the number of AGB
stars near the TRGB.  Moreover, all are sensitive to low-number 
statistics (Madore \& Freedman 1995).

Our goal is to test the sensitivity of the TRGB-distance method to 
variations in SFH rather than compare the different TRGB measurement
techniques.  Therefore, we simply adopt the technique of 
Sakai et al.\ (1997) because it is one of the most commonly used.  
This method involves applying the 
Gaussian-smoothed Sobel filter to the logarithmic
LF and weighting the filter output by the Poisson noise at each magnitude.  
This enhances large edges relative to small edges (which are assumed to be
noise).  The highest peak in the filter 
response within one magnitude of our
theoretically calibrated TRGB is taken to be the true TRGB.
To minimize spurious detections due to low-number statistics, we require
that at least 1000 stars be present within $\pm 1$ mag of the TRGB in
each of our synthetic CMDs.  This requirement applies to stars of 
all ages regardless of SFH so a purely old population will have 
no stars brighter than 
the TRGB whereas an intermediate-age population 
will have many stars brighter than the TRGB.
This requirement also means that we must normalize the 
overall SFRs to high levels not found in the LG dwarf galaxies.  
However, this is inconsequential to the 
current study because we are concerned with the {\it relative} SFRs at 
various ages and metallicities.  Small numbers of stars near the tip 
or any of the other problems mentioned above are beyond the scope 
of this work and have been examined in other papers 
(Madore \& Freedman 1995, Frayn \& Gilmore 2003).

To estimate the random error of $\MITRGB$, we follow the 
common technique of using the FWHM of a
Gaussian fit to the peak in the edge detector response 
(Sakai et al.\ 1999).  The total 
error of each of our TRGB estimates is then given by,
\begin{equation}
\sigma_{tot}^2 = \sigma_{theory}^2 + \sigma_{rand}^2,
\end{equation}
where $\sigma_{theory}$ is the standard deviation of $\MITRGB$ 
from the isochrones (0.05; see Fig. 2), 
and $\sigma_{rand}$ is the random error.  For all models,
$\sigma_{rand}$ ranges from $0.03-0.09$ and $\sigma_{tot}$
ranges from $0.06-0.11$.

\section{The Effect of Age Variations on $\MITRGB$}

For each of the models discussed below, we plot the relative SFR 
normalized to the average background rate, the 
resulting synthetic CMD containing 
10\% of the stars, logarithmic LF, and edge detector output.  
The dashed line marks the position of the estimated TRGB value 
while the dotted line marks our theoretical calibration of $-4.0$.

\subsection{Delta-function bursts}

We begin by simulating single-age, single-metallicity 
bursts at 2, 1.78, 1.58, and 
1.41 Gyr each with [Fe/H] = $-1.3$ (Fig.\ 6).
Even when observational effects are taken into account, the TRGB 
becomes fainter as the age decreases.  Since there is only one 
age in each case, the edge detector has no trouble accurately picking out 
the TRGB as can be seen by comparing the results with the theoretical
isochrones of Fig.\ 5.  Each TRGB estimate is well defined and unambiguous.

It is worth noting the gap between the TRGB and the AGB.  It is evident
as a lack of stars in the CMDs and a depression in the LF just above the 
TRGB.  This gap is important because its presence makes it easier for the
edge detector to find the TRGB by reducing the 
background level of AGB stars.  
The gap is not an artifact of the method but rather it is intrinsic to
the isochrones which predict very rapid evolution on this part of the AGB.
Interestingly, such a gap has been observed in Sextans B (Fig.\ 6 of M$\rm 
\acute{e}ndez$ et al.\ 2002) and the Magellanic Clouds 
(Fig.\ 5 of Cioni et al.\ 2000).

\subsection{Single Gaussian bursts}

It is possible that a mix of ages could hide the effect described in the
previous subsection.  To investigate this we generated Gaussian bursts in
age centered at the same ages as before, each with standard 
deviation of 0.2
Gyr and [Fe/H] = $-1.3$ (Fig.\ 7).  The edge detector is most 
sensitive to the largest discontinuity which is due
to the dominant population, in this case the center of each Gaussian.
In column 2 of Fig.\ 7, the presence of older stars pulls the derived TRGB 
up to a 
brighter magnitude than the equivalent delta-function burst.  So we can
expect that as the number of intermediate- and old-age stars increases, the
number of young stars needed to significantly change the
value of $\MITRGB$ also increases.
The resulting LFs are smoother and because there are 
several different ``competing'' TRGBs (one from each age), the 
discontinuity at each is 
smaller.  Consequently, the peak in the edge detector output corresponding
to the TRGB is not as prominent.

\subsection{Multiple Gaussian bursts}

We now add another level of complexity by simulating three background
Gaussian bursts
centered at 14.12, 7.08, and 1.58 Gyr.  The oldest burst has a standard
deviation of 2 Gyr and [Fe/H] = $-1.7$, the intermediate burst has a standard
deviation of 2 Gyr and [Fe/H] = $-1.4$, and the young burst has a standard
deviation of 1 Gyr and [Fe/H] = $-1.3$.  The SFR is zero outside the range 
$0.09-14.96$ Gyr.  First, each burst has an equal 
height as depicted in Fig.\ 8 (column 1).  Then we add a short,
quick burst by
scaling the SFR between 1.33 and 1.68 Gyr to 
$\sim 12$ times (Fig.\ 8, column 2), 
$\sim 25$ times (Fig.\ 8, column 3), and 
$\sim 35$ times (Fig.\ 8, column 4)
the average background rate.  This young burst is timed so that
the stars it forms are undergoing the RGB phase
transition at the present time.

In the first and second cases, the TRGB is unambiguously detected very close
to $\MI = -4.0$.  In the third case, 
when the young, quick burst is scaled to $\sim 25$ times the background,
the edge detector responds almost equally to the TRGB for the quick 
burst (at $\MI \sim -3.7$) and the 
TRGB for the older stars.  
In the fourth case, the edge detector responds primarily to the TRGB 
of the young population although the TRGB of the old population is 
still visible at $\MI = -4.0$.  In a real stellar population this
could easily be mistaken for noise due to AGB stars.

\subsection{Constant star formation rate}

Next we consider a constant SFR from 14.96 to 0.09 Gyr ago.
To simulate the metallicity evolution of an LG dwarf galaxy, we use two
different chemical enrichment laws (CELs).  These laws are plotted in 
Fig.\ 9 and described below.

\subsubsection{Chemical enrichment law \#1}

Our first CEL is adapted from Aparicio et al.\ (1996) for our age range.  The 
metal abundance
increases from [Fe/H] = $-1.7$ to $-0.7$.  Because our set of isochrones has
discrete values of metallicity we approximate the smooth 
change of [Fe/H] with time by arranging the isochrones 
into age groups each with constant metallicity.

First, we use a constant SFR (Fig.\ 10, column 1).  Here the TRGB
is recovered accurately and precisely.  Then we introduce
a burst between 1.33 and 1.68 Gyr by scaling the SFR at these ages to 20
times (Fig.\ 10, column 2) and 30 times (Fig.\ 10, column 3) 
the background rate.  In both cases, the younger, fainter TRGB is 
comparable in size to the older, brighter TRGB.  Hence, the edge detector 
responds roughly equally to the TRGBs of the young and old populations 
but the young TRGB is slightly more dominant.  This highlights the
importance of careful examination of the edge detector output in the
vicinity of the TRGB.  If there are peaks of similar
size in the output then the random error may be underestimated 
by simply fitting a Gaussian to the highest peak.  Lastly, we
scale the young burst up to 40 times the background rate 
(Fig.\ 10, column 4).  Here the 
younger, fainter TRGB dominates the edge detector response.

\subsubsection{Chemical enrichment law \#2}

We repeat the procedure from the previous subsection using a different
CEL that describes a more gradual metal enrichment (Fig.\ 9).  
The only change is that the
final metal abundance is decreased to [Fe/H] = $-1.3$.  
Since the average metallicity of this population is lower than 
before, the RGB is steeper and the upper AGB is shorter.  
As before, we start
with a constant SFR (Fig.\ 11, column 1) and add a successively 
stronger burst between 1.33 and 1.68 Gyr.  
When the young burst is 20 times (Fig.\ 11, column 2) and 30 times
(Fig.\ 11, column 3) the background rate,
the edge detector appears stable and accurate.  When the young 
burst is 40 times the background rate (Fig.\ 11, column 4), 
the edge detector output displays several peaks of comparable
height but the highest is due to the TRGB of the young burst.

\subsubsection{No intermediate-age stars}

If there were no intermediate-age stars, it should be easier for a young
burst of star formation to move $\MITRGB$ to fainter magnitudes because there
would be fewer stars populating the TRGB at $\rm M_I = -4.0$.  To check this
we now investigate what happens when there is a constant SFR from 
$6.68-14.96$ Gyr and $1.68-1.33$ Gyr with zero SFR everywhere else.  
The metallicity evolves from [Fe/H] = $-1.7$ for ages
$>$ 10.59 Gyr to $-1.4$ for ages between 6.68 Gyr and 10.59 Gyr
and then to $-0.9$ for the young burst.  
We show in Fig.\ 12 that the TRGB is detected accurately and precisely
when the young burst is equal to and 10 times larger than the old SFR.
But when the young burst is 15 and 20 times larger, 
the edge detector responds only to the TRGB of the young burst.

Finally, we repeat the SFH from above except that the SFR between 6.68
Gyr and 10.59 Gyr is set to zero.  When the young burst has the same
SFR as the old burst the TRGB of the old burst is recovered correctly 
(Fig.\ 13, column 1).
When the young SFR is scaled to five times the old SFR the discontinuity
due to the young TRGB is slightly more significant than that due to the
old TRGB (Fig.\ 13, column 2).  The young TRGB dominates the edge 
detector output when the young SFR is
10 times (Fig.\ 13, column 3) and 15 times (Fig.\ 13, column 4) 
the old SFR.

\section{The Effect of Metallicity Variations on $\MITRGB$}

\subsection{Delta-function bursts}

As before, we begin with single-metallicity, single-age bursts but we now 
vary the metallicity and hold the age constant at 10.00 Gyr.  This age
was chosen to ensure that the effects described in the previous section
would not interfere with the results of this section.
The burst in Fig.\ 14 (column 1) has a metallicity of [Fe/H] = $-0.7$.  
For this case, the largest peak in the edge detector response
lies at $\rm M_I \sim -4.7$.  This peak must be coincident with the 
AGB tip rather than the TRGB otherwise
there would be an unphysically large break in the power-law 
slope of the RGB LF at $\rm M_I \sim -4.0$.  Hence, the second-largest
peak in the edge detector response which lies at $\rm M_I = -4.05 \pm 0.06$ 
must correspond to the TRGB; a fact which we are able to confirm
because the masses of the synthetic stars are known.

When the metal abundance is increased to [Fe/H] = $-0.4$ 
(Fig.\ 14, column 2), $\MITRGB$ is still close to the
canonical value but has moved 0.1 mag fainter.
Moreover, the slope of the RGB in the CMD has
become so shallow that the tip of the AGB occurs at nearly the 
same magnitude as the TRGB.  In the logarithmic LF, the TRGB has 
almost merged with the tip of the AGB which shows a slightly
less significant discontinuity.

In Fig.\ 14 (column 3), the burst has solar metal abundance.  
Here there are no AGB stars past the TRGB so the final drop-off in 
the logarthmic LF is the TRGB rather than the AGB tip.
If one
were to assume the metallicity of this population was less than $-0.7$,
the TRGB brightness would be underestimated by 0.76 mag.
An even greater underestimate would occur when [Fe/H] = 0.2 
(Fig.\ 14, column 4), for
which the recovered TRGB is over a full magnitude fainter than the 
canonical value.  Note that in these two cases, we had to focus on the
magnitude interval $-3.0 \pm 1.0$ rather than $-4.0 \pm 1.0$ 
because the slope of the RGB was so shallow.

The four cases above illustrate that the RGB becomes flatter
as metallicity increases.  This is a well-known 
phenomenon caused by line blanketing primarily in the V-band but also
in the I-band due to the increased metal abundance.

\subsection{Single Gaussian metallicity distribution}

To investigate the effect of a spread in metallicities we simulated
four Gaussian bursts in metallicity each centered on the same metallicities
and with the same age as the delta-function bursts.  The dispersion of each
burst was 0.2 dex.  Because our isochrones cover a discrete set of 
metallicities we can only sample each burst at these metallicities.  This
results in each CMD having several clearly defined RGBs for each metallicity.

When the Gaussian distribution of metallicities is centered on 
[Fe/H] = $-0.7$ and $-0.4$, $\MITRGB$ lies 
within 1$\sigma$ of the canonical value 
(Fig.\ 15, columns 1 and 2).  
But when the peak metallicity is [Fe/H] = 0.0 and 0.2 
(Fig.\ 15, columns 3 and 4),
the estimated TRGB is about 0.76 mag fainter than the canonical value.
In the last case, the TRGB is recovered 0.4 mag
brighter than the equivalent delta-function burst.
This may be due to the presence of more metal-poor stars populating
the TRGB at brighter magnitudes.

Next, we set the SFR in each metallicity bin to be constant from ages
$9.44-14.96$ Gyr (Fig.\ 16).  This has the effect of smearing 
out the RGB of each metallicity.  The results are identical except 
that the recovered TRGB is about 0.1 mag fainter for peak 
metallicities of [Fe/H] = 0.0 and 0.2.

\subsection{Uniform metallicity distribution}

Next we examine the case of a uniform metallicity distribution 
and constant SFR from $9.44-14.96$ Gyr in the past (Fig.\ 17, column 1).
The random error is small and the canonical value lies within the total
uncertainty of $\MITRGB$.  
The result is the same when we increase the
SFR at solar abundance to five times the background value 
(Fig.\ 17, column 2).  But at 10 and 20 times 
the background value (Fig.\ 17, columns 3 and 4), 
the edge detector is biased toward the TRGB of this burst.
Note that in these latter two cases, the canoncial TRGB due to 
lower metallicities is still somewhat visible in the edge
detector output.

Using the same uniform metallicity distribution and SFR, we repeated
the previous procedure but lowered the metallicity of the burst 
to [Fe/H] = $-0.1$ (Fig.\ 18).  When the burst is five and 10
times stronger than the background rate, the TRGB is 
close to the canonical value.  But when the burst is
20 times the background, $\MITRGB$ moves 0.35 mag
fainter than the canonical value.

Repeating the procedure with a slightly different metallicity
distribution which is uniform but has no stars with 
[Fe/H] $< -0.9$, the canonical TRGB
is recovered to within the total error (Fig.\ 19, column 1).  
If the
SFR at solar abundance is again increased by a factor of five 
(Fig.\ 19, column 2) then the
discontinuity due to the TRGB of the burst is larger than
that due to the canonical TRGB.  
When the burst is increased to 10 times (Fig.\ 19, column 3) 
and 20 times (Fig.\ 19, column 4) the background rate 
the edge detector response is unambiguously biased to fainter 
magnitudes.

\subsection{Multiple Gaussian metallicity distribution}

To better simulate the metallicity distribution of real
galaxies, we next introduce
three Gaussian bursts in metallicity.
For the parameters of each Gaussian, we use the metallicity distribution 
function of Sarajedini \& Van Duyne (2001) for M31.  
The metal-poor component has 
$\rm [Fe/H] = -1.50 \pm 0.45$, the intermediate-metallicity component
has $\rm [Fe/H] = -0.82 \pm 0.20$, and the metal-rich component has
$\rm [Fe/H] = -0.22 \pm 0.26$.  The SFR at each
metallicity is constant and non-zero over the ranges $9.44-14.96$, 
$5.96-10.59$, and $2.11-6.68$ Gyr for the metal-poor, -intermediate, 
and -rich
components, respectively.  Each component is normalized to have equal
area but because of our finite metallicity coverage, the metal-rich and
metal-poor components are truncated.

In this case (Fig.\ 20, column 1) the dominant population
observed today has $\rm [Fe/H] \sim -0.8$ and age $\sim$ 6 Gyr 
so $\MITRGB$ is $1\sigma$ brighter than the canonical value.
But when we introduce a burst at [Fe/H] = 0.0 by scaling the 
SFR up to $\sim 8$ times larger than the average background rate
(Fig.\ 20, column 2), the edge detector output is ambiguous 
because it is dominated by several different peaks of 
nearly equal height.  
When the same metal-rich burst is scaled to $\sim 15$ times 
(Fig.\ 20, column 3) 
and $\sim 25$ times (Fig.\ 20, column 4) the average 
background rate, the 
edge detector is biased toward fainter magnitudes.

\section{Results}

As a quantitative measure of the importance of young stars, we 
introduce the ratio, $R$, of the number of stars in the RGB phase 
transition stage within $\pm 1$ mag of the measured TRGB to the total 
number of stars in the same magnitude range.
%\begin{equation}
%$R \equiv \frac{N_{pt}}{N_{tot}}$,
%\end{equation}
%where $\rm N_{pt}$ is the number of stars in the RGB phase 
%transition stage within one magnitude of the TRGB and 
%$\rm N_{tot}$ is the total number of stars in the 
%same magnitude range.  
In Fig.\ 21 we plot the measured distance modulus assuming 
$\MITRGB = -4.0$ as a function of R.  If the young population is 
small, then $\rm (m-M)_I$ is correctly recovered at 0.0.  
When R $\sim$ 0.50, the young population is half the 
total population near $\MI = -4.0$ and the 
two competing discontinuities are relatively 
comparable in size.  At this point the edge detector responds to both
roughly equally.
For R $>$ 0.60, the
young population dominates the LF and the TRGB method is 
inaccurate, leading to overestimates of $\rm (m-M)_I$ by 
$0.2-0.5$ mag.

The quantity, R, is not observable in practice.  We would like to use another
quantity that provides the same information about the accuracy of the 
TRGB-distance method but is directly observable.  For
this purpose, we use the height of the discontinuity at the TRGB, $c$.  For 
each model we make separate linear fits to the logarithmic LF in
the regions one magnitude fainter and brighter than the measured TRGB.  Then
$c$ is simply the difference between the value of each line at the TRGB.
This definition of $c$ is similar to that of $\rm M\acute{e}ndez$ et al.\ 
(2002), the only difference being ours involves a fit to the observed LF 
while theirs involves a fit to the intrinsic LF.

We would expect that as the number of stars undergoing the RGB
phase transition increases
the TRGB discontinuity of the older population at $\rm M_I = -4.0$ 
eventually disappears in the AGB of the young population.  
Moreover, the TRGB of the young population is diminished by the RGB of
the older population.  Hence, for a mix of ages, $c$ should decrease as 
R increases.  In other words, the LF in the vicinity of $\rm M_I = -4.0$
should get smoother as R increases.
This is demonstrated in Fig.\ 22 which shows the LFs of the
models from \S 3.3.
The same relationship cannot be true for purely young
populations because they contain no stars with discordant 
values of $\MITRGB$ to smooth their LFs.

Such is the case in Fig.\ 23 where we plot $c$ as a function of R.  
In this plot, the error bars are the random errors of the fit parameters
added in quadrature.  The
purely young models (R $>$ 0.9) do not follow any relationship whereas the
models with a mix of ages show an inverse relationship like the one we would 
expect.  To check the strength of the correlation we compute Kendall's Tau
rank correlation coefficient, $\tau$, for all the models and then for 
just the models with R $<$ 0.9.  In the former case, 
$\tau = -0.18$ indicating a negative correlation exists
at the 91\% confidence level.  In the 
latter case, $\tau = -0.52$, 
indicating a negative correlation exists at the 99.96\% confidence level.  
By inspection, values of $c \lesssim$ 0.4 correspond to R $\gtrsim$ 0.55.  
Hence, values of $c \lesssim$ 0.4 should give errant 
measurements of $\rm (m-M)_I$.

This expectation is confirmed in Fig.\ 24 which shows our estimated distance 
modulus as a function of $c$.  Models with $c \lesssim$ 0.4 do indeed 
overestimate $\rm (m-M)_I$ by
about $0.3-0.5$ mag corresponding to a fractional distance error 
of $\sim 15-25\%$.  
Note that we have lost some information in the transition from R to $c$
because values of $c \gtrsim$ 0.4 do not guarantee an accurate 
estimate of the 
TRGB.  However, $c$ is the more useful parameter because it
is directly available from the LF.  Finally, large photometric errors
at the magnitude of the TRGB could cause a similar smoothing
effect to that discussed above.

As another diagnostic, we define the critical SFR, $\rm SFR_{crit}$, 
as the SFR of a burst between 1 and 2 Gyr ago 
which causes the TRGB-distance method to fail.
This critical SFR is measured relative to the average background rate
and depends on the duration of star formation at older ages.
When the background star formation occurs over the lifetime of the
Universe, $\rm SFR_{crit} \sim 20-40$ (Figs.\ 8, 10, and 11).  
If the background star formation ended about 7 Gyr ago,
then $\rm SFR_{crit} = 15$ (Fig.\ 12) but $\rm SFR_{crit} = 5$ for a 
background that ended about 10 Gyr ago (Fig.\ 13).  If there were little
or no stars created at older ages then $\rm SFR_{crit} \sim 1-4$
(Figs.\ 6 and 7).
When the SFR between 
1 and 2 Gyr ago equals $\rm SFR_{crit}$, the distance 
modulus could be overestimated by $\sim 0.2-0.5$ mag.

In addition to the duration of star formation at old ages, 
$\rm SFR_{crit}$ also depends on the duration of the burst
between 1 and 2 Gyr ago.  For example, if the burst
was 0.1 Gyr rather than 0.4 Gyr long, 
$\rm SFR_{crit}$ would be larger.  To account for varying strengths and
durations of star formation at old and young ages, we define
the ratio, $W$, of the 
number of stars formed $1-2$ Gyr ago to the
total number of stars formed over the lifetime of the Universe.
In contrast to $R$, $W$ measures the
number of stars formed $1-2$ Gyr ago rather than the number 
of such stars that are observable
today in a particular location of the CMD.
In Fig.\ 25, we plot the measured distance modulus versus $W$.
When $W > 0.30$ there
is a 78\% chance of overestimating the distance modulus.

We can understand the effect of metal-rich stars in a similar manner by
defining the ratio, $Q$, of the number of metal-rich stars
(i.e. [Fe/H] $> -0.7$) located within $\pm 1$ mag of the 
measured TRGB to 
the total number of stars in the same magnitude range.  
%\begin{equation}
%$Q \equiv \frac{N_{mr}}{N_{tot}}$,
%\end{equation}
%where $\rm N_{mr}$ is the number of metal-rich stars defined as having
%[Fe/H] $> -0.7$ and located within one magnitude of the TRGB and 
%$\rm N_{tot}$ is the total number of stars in the same magnitude
%range.  
In Fig.\ 26, we plot the distance modulus assuming 
$\MITRGB = -4.0$ as a function of Q.  The method is accurate up to 
Q $\sim$ 0.6
at which point significantly errant measurements begin to occur.  
In fact, 78\% of the 
measurements with Q $>$ 0.6 are overestimated by $> 0.2$ mag.

We can also directly explore how the 
metallicity of stars affects the TRGB distance method.  This is shown
in Fig.\ 27 where $\rm \langle[Fe/H]\rangle$ is defined as the average 
metallicity of stars
in the range $0.4-0.6$ mag fainter than the estimated $\MITRGB$ for each 
model.  This magnitude range corresponds to $\rm M_I = -3.5 \pm 0.1$ 
in the absence of any other information about the true distance modulus.
Several models from \S 3 (column 1 of Figs.\ $6-8$ and $10-13$) 
are plotted as open circles to extend the coverage to 
lower metallicities.  The estimated distance modulus 
is accurate up to 
$\rm \langle[Fe/H]\rangle \sim -0.3$.  Above this metallicity the 
distance modulus is overestimated by about $0.2-1.0$ mag.

It is more useful to relate the error in distance modulus to a directly
observable quantity.  Hence, we plot in Fig.\ 28 the 
median $\rm (V-I)_0$ versus $\rm \langle[Fe/H]\rangle$
for stars $0.4-0.6$ mag fainter than $\MITRGB$.  
Again, the open circles represent several 
models from \S 3.
As expected, this confirms the well known relation 
between the color of the TRGB and metallicity.  This graph
shows that $\rm \langle[Fe/H]\rangle > -0.3$ corresponds
to median $\rm (V-I)_0 \gtrsim$ 2.0.  So we would expect that
the models with median $\rm (V-I)_0 \gtrsim$ 2.0 have values
of $\MITRGB$ significantly different from the canonical value.

This expectation is confirmed in Fig.\ 29 which shows
the distance modulus as a function of the median $\rm (V-I)_0$.  
The distance moduli of all models 
with median $\rm (V-I)_0 < 1.9$ were recovered accurately but 67\% of the 
models with median $\rm (V-I)_0 > 1.9$ and 86\% with median
$\rm (V-I)_0 > 2.0$ yielded distance moduli too large
by about $0.2-1.0$ mag.

Now we define $\rm SFR_{crit}$ as the SFR 
of a burst with [Fe/H] $> -0.3$ required to cause the TRGB-distance 
method to fail.
This critical SFR is measured relative to the average background 
rate at lower metallicities and depends on the range of low
metallicities present in the metallicity distribution.  
When the background star formation 
occurs for $\rm -1.7 \leq [Fe/H] \leq -0.3$, then
$\rm SFR_{crit} \sim 10-20$ (Figs.\ 17, 18, and 20).  
If the background star formation occurs over a less extended
range of metallicities, like 
$\rm -0.9 \leq [Fe/H] \leq -0.3$, then $\rm SFR_{crit} = 5$ (Fig.\ 19).
If there were little or no stars created at low metallicities
then $\rm SFR_{crit} \sim 1-2$ (Figs.\ $14-16$). 
When the SFR at [Fe/H] $> -0.3$ equals $\rm SFR_{crit}$, 
there is aproximately a 78\% chance that the distance modulus
will be overestimated by 
$\sim 0.2-1.0$ mag.
This assumes that the SFR at each metallicity is constant
over approximately the same amount of time.  If the star formation for
[Fe/H] $< -0.3$ occurs over a shorter time span, then 
$\rm SFR_{crit}$ would be reduced.

In addition to the range of low metallicities created, 
$\rm SFR_{crit}$ also depends on the range of high 
metallicites created.  For example, if the burst at high
metallicities was 0.1 dex rather than 0.2 dex wide, 
$\rm SFR_{crit}$ would be larger.  To account for varying
strengths and widths of metal-poor and -rich bursts, we define
the ratio, $X$, of the
number of stars created with [Fe/H] $> -0.3$ to the
total number of stars created at all metallicities.
Unlike $Q$, $X$ measures the number of metal-rich stars
formed rather than the number of such stars observable today
in a particular location in the CMD.
In Fig.\ 30, we plot the measured distance modulus versus $X$.
When $X > 0.70$ there
is a 71\% chance of overestimating the distance modulus.

Tables 1 and 2 summarize the results of all simulations.  The last column
in both tables is the fractional distance error.

\section{Summary and Conclusions}

We have shown that the TRGB-distance method is insensitive to star formation
history except for large bursts between ages of about 1 and 2 Gyr. 
Stars formed at these ages are today undergoing the RGB phase
transition during which time the TRGB lies at magnitudes
fainter than the canonical value.
Stars less than 1 Gyr old reach the TRGB at a significantly bluer
location in the $\rm M_I - (V-I)_0$ CMD than stars greater than 
2 Gyr old.  These 
very young stars should not significantly affect the TRGB-distance method
because they simply add background noise which could be minimized by 
excluding them from the analysis as is often already done
(Jerjen \& Rejkuba 2001; Ma$\rm \acute{i}$z-Apell$\rm \acute{a}$niz
et al.\ 2002; Sakai et al.\ 2000; Sakai et al.\ 1999; 
M$\rm \acute{e}$ndez et al.\ 2002).
The RGB length is so short for $1-1.3$ Gyr that one could detect such a
population, if the photometry reveals the RC, by measuring the full
length of the RGB between the TRGB and RC.  But for the 
middle and late stages of the
RGB phase transition ($\sim 1.3-1.7$ Gyr), observational errors might
prohibit the use of the RGB length as an indicator.  

Instead, the size of the 
discontinuity, $c$, in the LF at the TRGB can be used as a rough 
indicator of a large burst between $\sim 1.3-1.7$ Gyr provided that at 
least 10\% of the population was formed well over 2 Gyr ago.  
If $c \lesssim$ 0.4 then there may be a 
significant population undergoing the RGB phase transition
which comprises at 
least $\sim$ 60\% of the total number of stars
within one magnitude of the TRGB.  These stars would dominate 
the LF and the assumption that $\MITRGB \approx -4.0$ would cause an 
overestimate of the distance modulus.

Equivalently, if the strength and duration of star formation 
between the ages of 1.3 and 1.7 Gyr are such that over 30\% 
of the total number of stars created are created in this age range, 
then there is a 78\% chance the
distance modulus will be overestimated by $\sim 0.2-0.5$ mag.  
This corresponds to a distance error of $\sim 10-25\%$.

When the number of stars with [Fe/H] $> -0.7$
is at least $\sim$ 60\% of the total number of stars within one magnitude
of the TRGB, there is a $\sim 78\%$ chance they will 
move $\MITRGB$ significantly fainter than the canonical value.
This means that when the average metallicity
of stars 0.5 mag fainter than the TRGB is $> -0.3$, 
or alternatively, when the median
$\rm (V-I)_0$ of these stars is greater than 1.9, $\MITRGB$ 
could be about $0.2-1.0$ mag fainter than $\rm M_I = -4.0$.
We caution that the RGBs of 
the isochrones we have used in this study tend to
be about 0.2 mag bluer than empirical RGBs.  Therefore, in practice
the critical color is probably closer to 2.1.

If the strength and duration of star formation at [Fe/H] $> -0.3$ 
are such that over 70\% of the total number of stars created have 
[Fe/H] $> -0.3$,
then there is aproximately a 71\% chance the distance modulus
will be overestimated by $\sim 0.2-1.0$ mag.  This corresponds 
to a distance error of $\sim 10-45\%$.

Two final notes of caution deserve to be mentioned again.  First, the
precise age and duration of the RGB phase transition depend
on the isochrones used.  In particular, models without convective
overshoot tend to predict a somewhat earlier age and shorter duration.
Second, the sensitivities of the TRGB to variations in SFH which
we have examined may change under error, completeness, and crowding 
conditions significantly different from those we have assumed.  This 
is an issue that deserves further study.

Nevertheless, it is hard to imagine how such strong bursts localized in 
such a small range of ages or metallicities could actually
occur in nature.  Indeed, no such bursts have been found in the 
SFHs of the LG dwarf galaxies (Harris \& Zaritsky 2001; 
Tolstoy et al.\ 1998; Gallart et al.\ 1996a;
Gallart et al.\ 1996b; Schulte-Ladbeck et al.\ 2002; Lynds et al.\ 1998; 
Gallart 2000; Carrera et al.\ 2002; Dolphin 2002).  We, therefore, conclude
that distance estimates using the TRGB method up to this time have not
fallen victim to the effects discussed in this paper.  However, as we
continue to resolve stellar populations that were previously too far to 
resolve, it is possible that nature will surprise us yet again.

\acknowledgements We thank the referee whose comments greatly
improved this paper.  We also thank Aaron Grocholski, Glenn Tiede for
many fruitful discussions, and Dennis Zaritsky for helpful 
comments on a draft.

\clearpage
\begin{figure}
\figurenum{1}
\epsscale{1.0}
\plotone{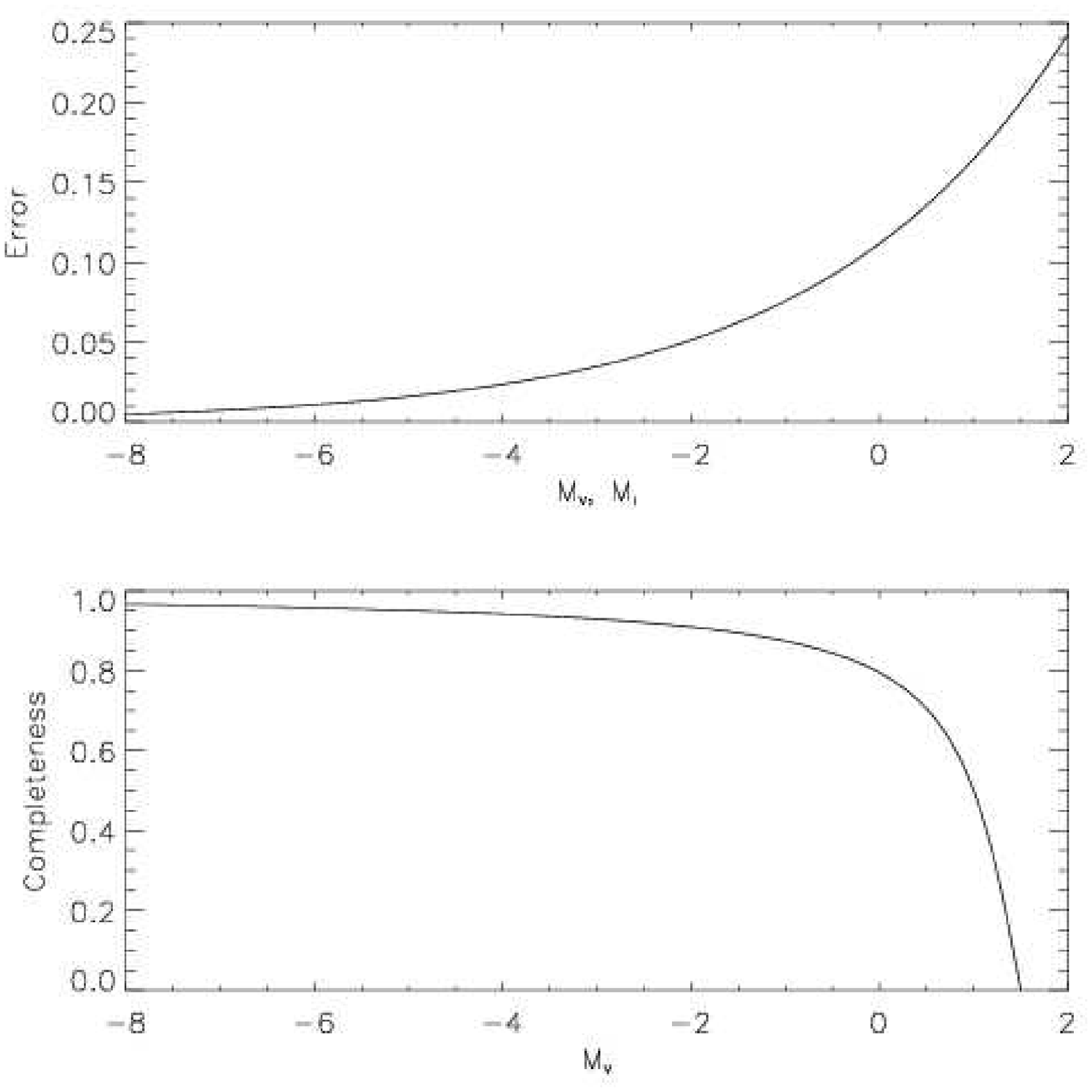}
\caption{Photometric error in $\rm M_V$ and $\rm M_I$ (upper panel) and 
completeness fraction as a function of $\rm M_V$ (lower panel).}
\end{figure}

\clearpage
\begin{figure}
\figurenum{2}
\epsscale{1.0}
\plotone{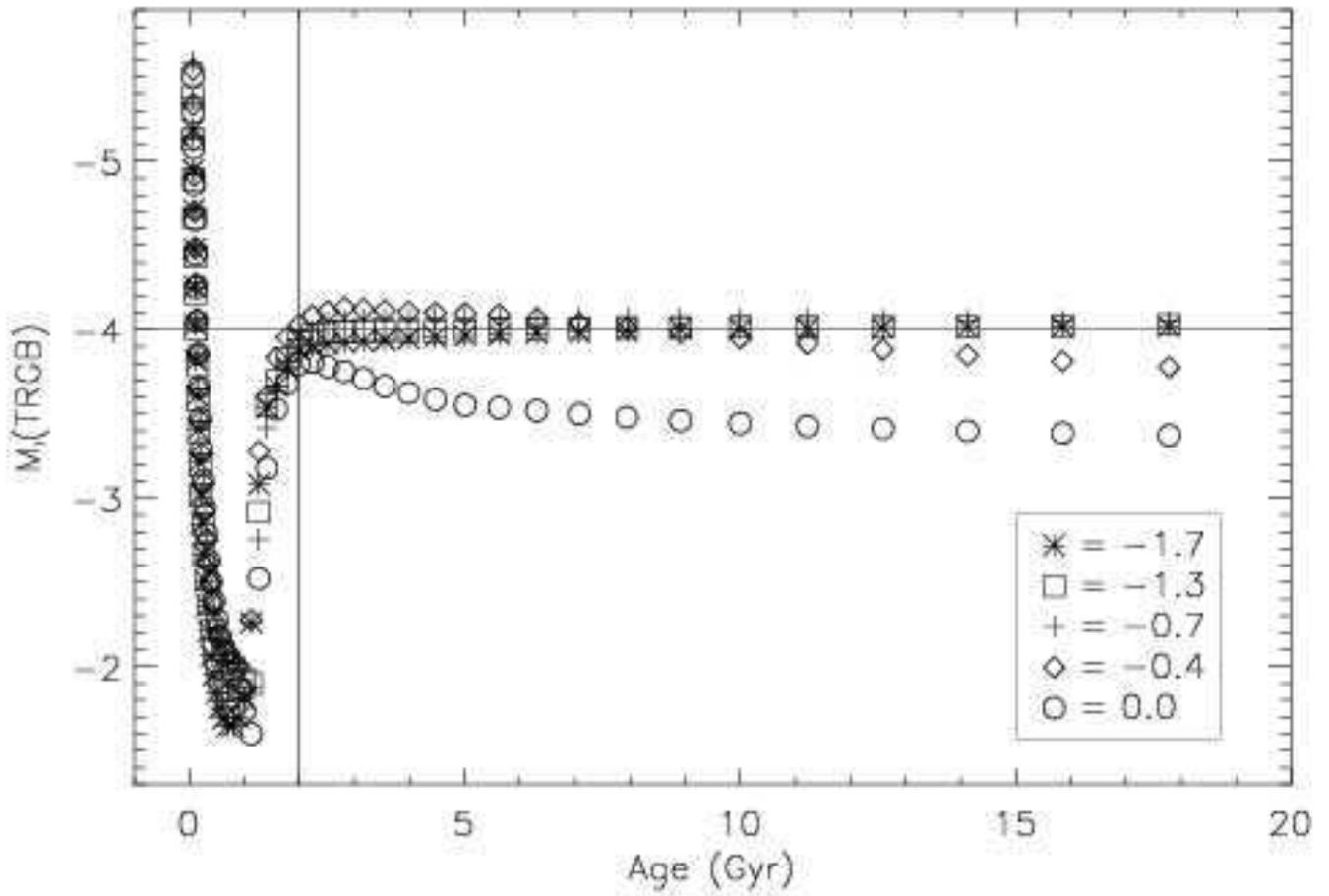}
\caption{I-band absolute magnitude of the TRGB as a 
function of age from the summary tables of the
Padova isochrones.  
[Fe/H] is given in legend at lower right.}
\end{figure}

\clearpage
\begin{figure}
\figurenum{3}
\epsscale{1.0}
\plotone{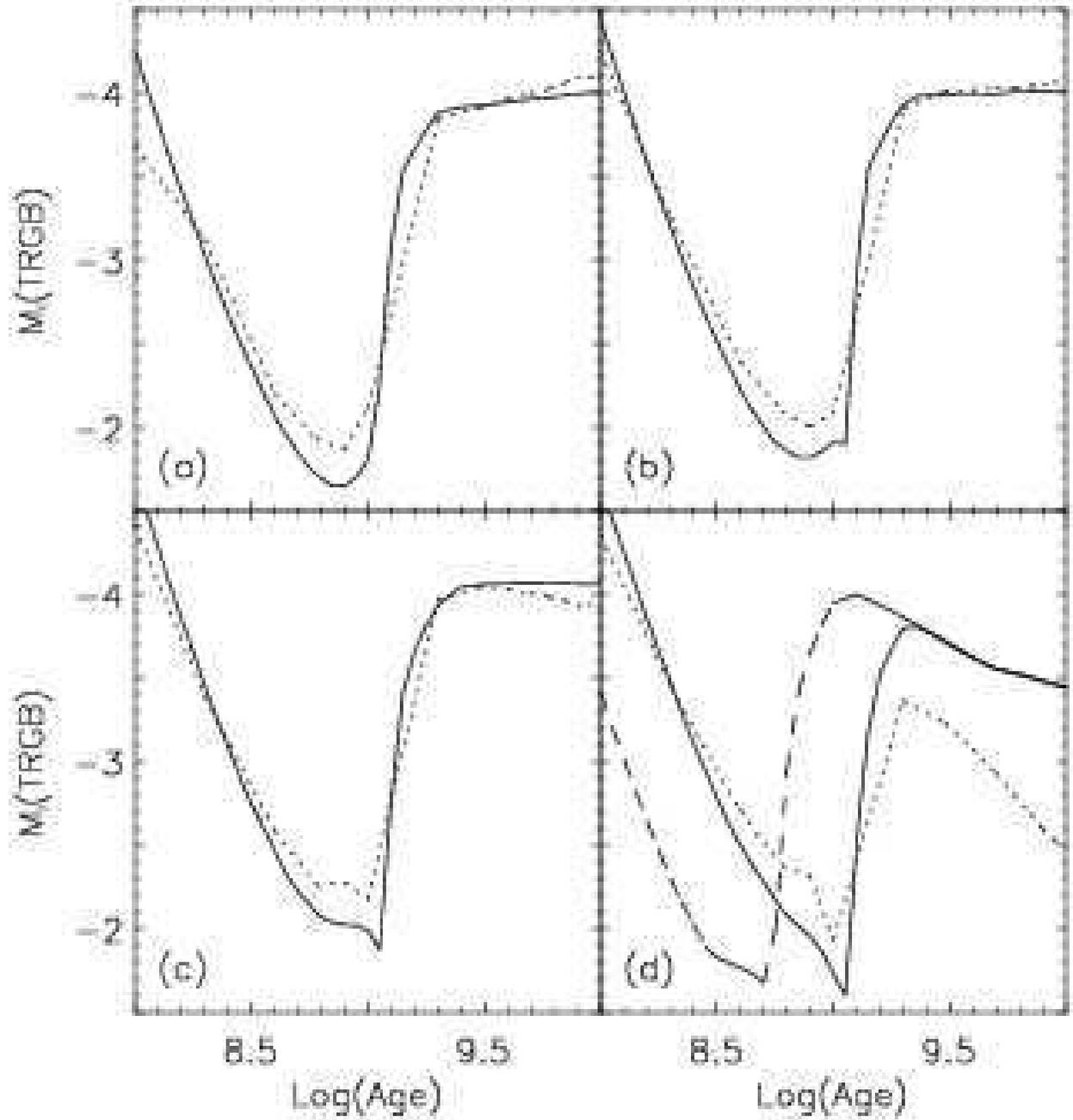}
\caption{Comparing the RGB phase transition of the Padova 
isochrones (solid) with that of the $\rm Y^2$ isochrones (dotted)
for metallicities of 
(a) [Fe/H] = -1.7, (b) -1.3, (c) -0.7, and
(d) 0.0.  The dashed curve in panel (d) corresponds
to the Padova isochrone with no convective overshoot.}
\end{figure}

\clearpage
\begin{figure}
\figurenum{4}
\epsscale{1.0}
\plotone{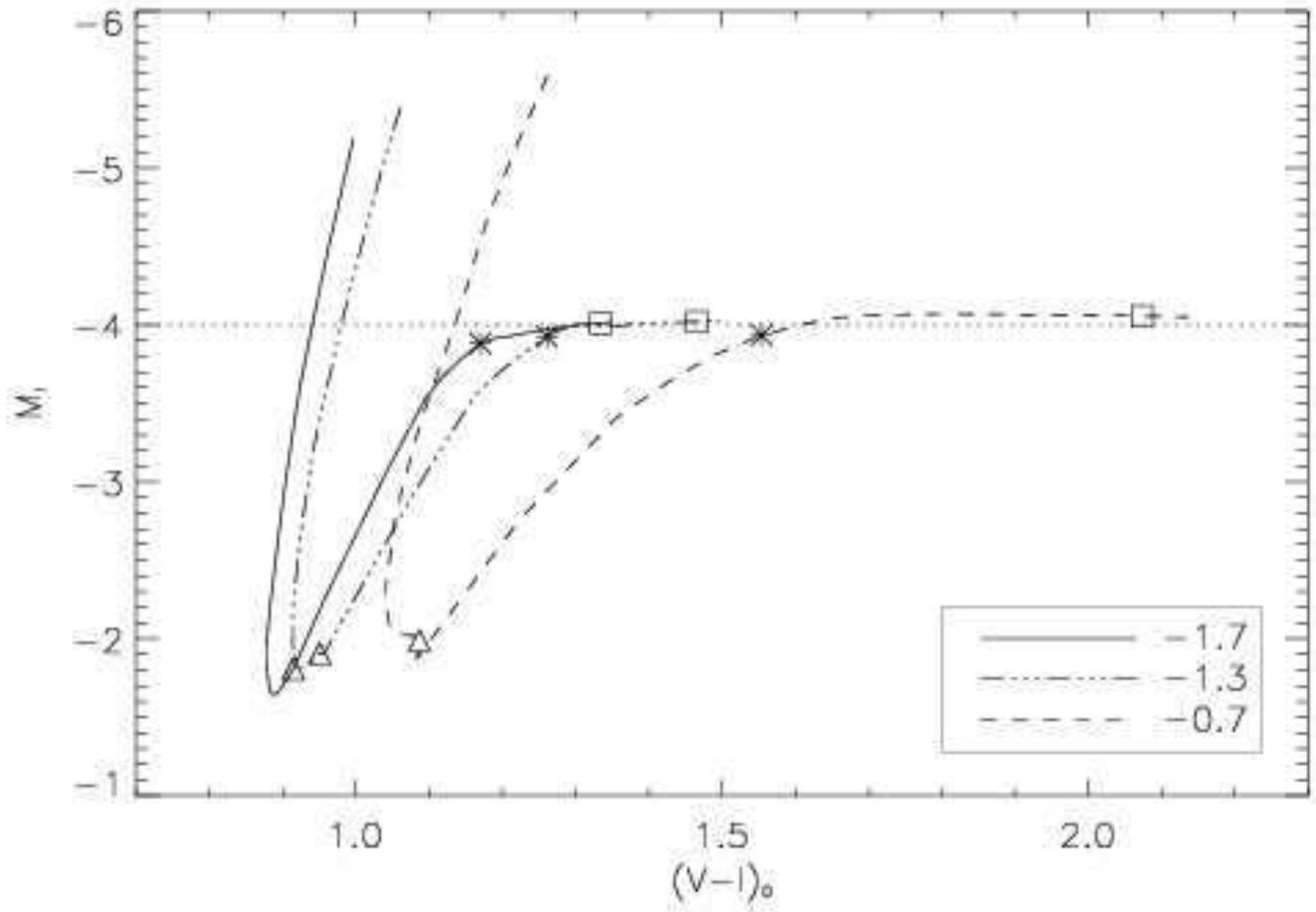}
\caption{Evolution of the TRGB in the $\rm M_I-(V-I)_0$ plane.  Triangles
mark 1 Gyr, asterisks mark 2 Gyr, and squares mark 14 Gyr.  [Fe/H] is 
given in legend at lower right.}
\end{figure}

\clearpage
\begin{figure}
\figurenum{5}
\epsscale{1.0}
\plotone{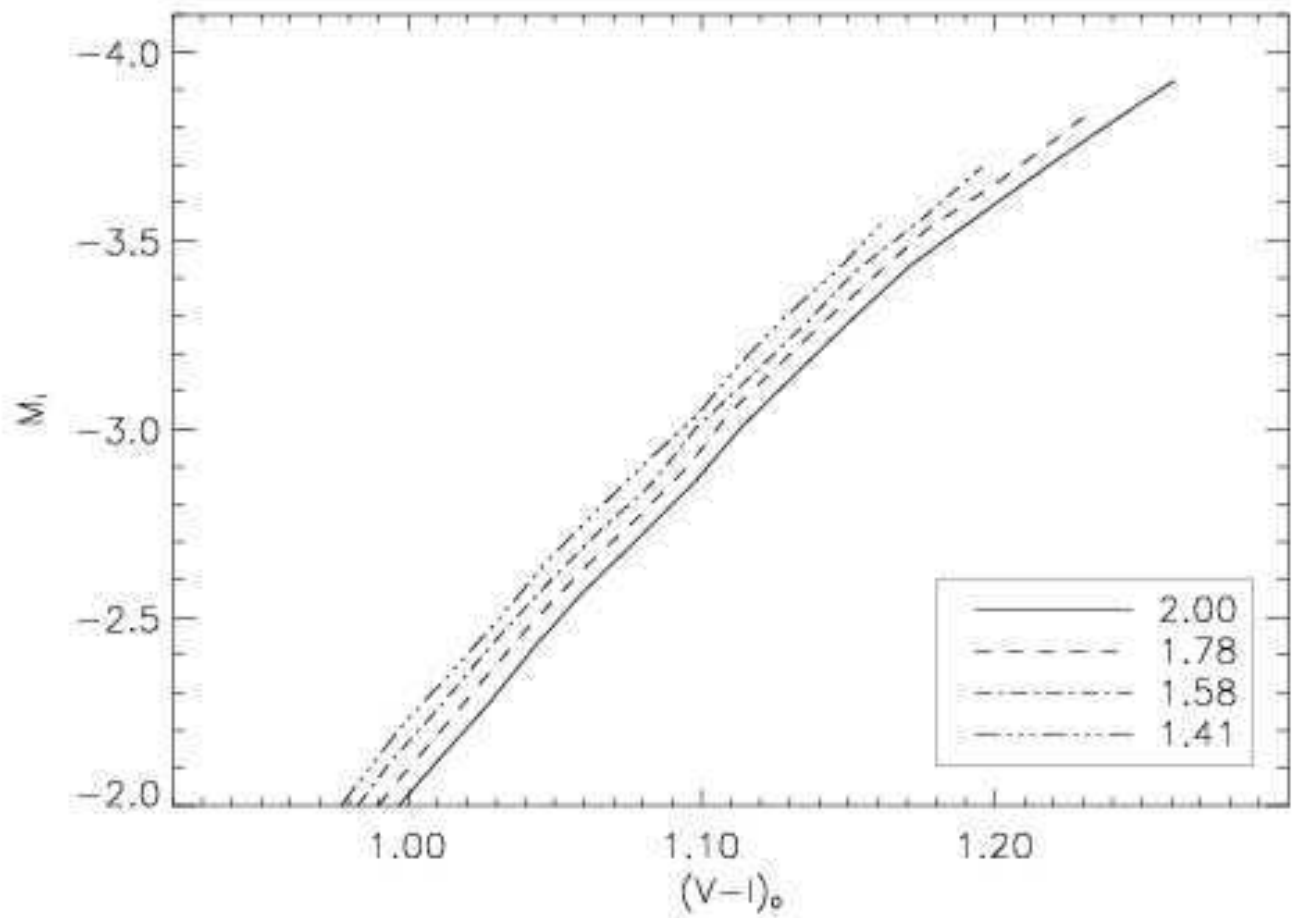}
\caption{RGBs during late stages of the RGB phase transition.  Ages in legend
are in Gyr and all have [Fe/H] = $-1.3$.}
\end{figure}

\clearpage
\begin{figure}
\figurenum{6}
\epsscale{1.0}
\plotone{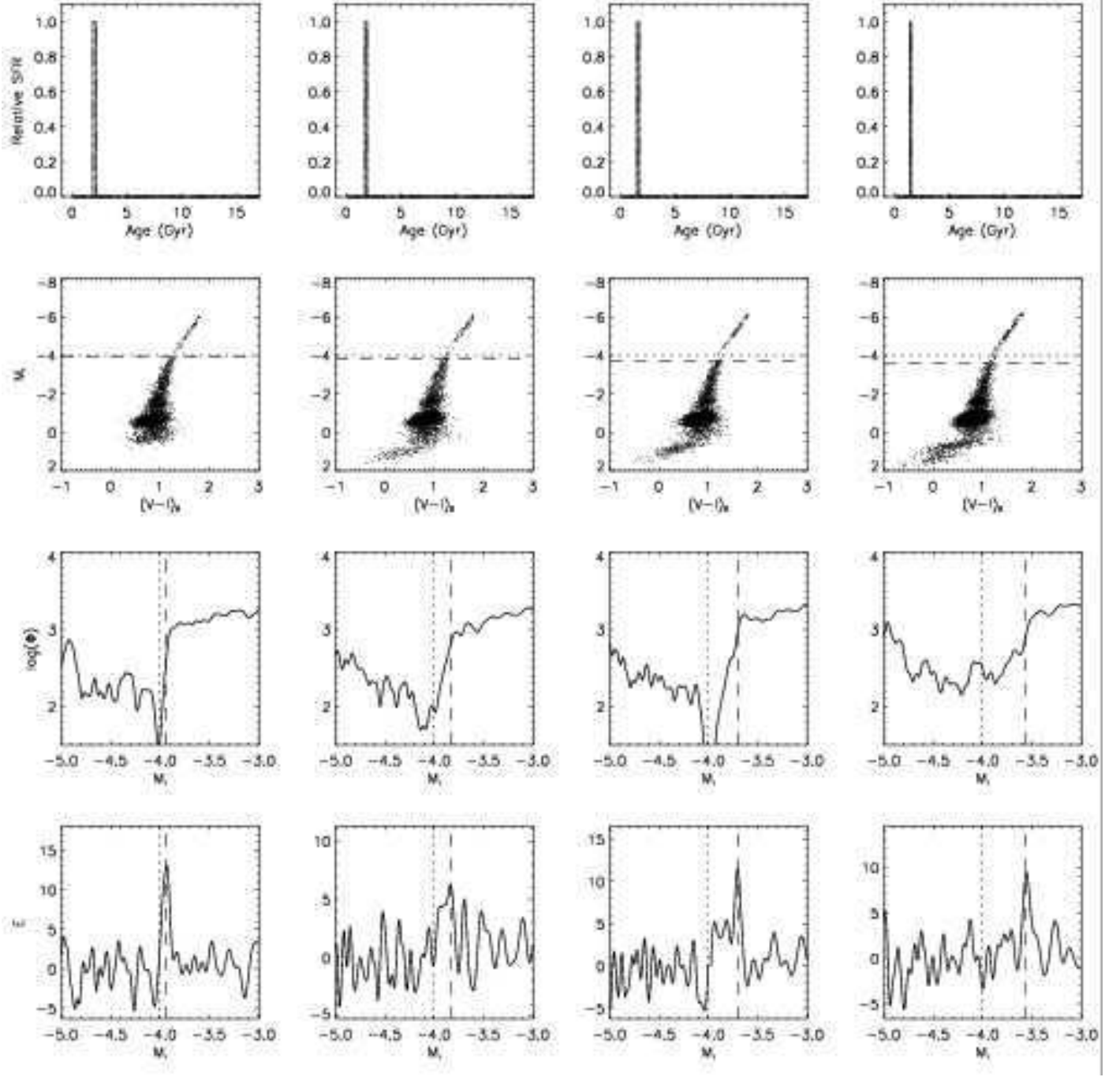}
\caption{From top to bottom:  SFR normalized to average
background rate, 
CMD containing 10\% of the stars, logarithmic LF, and
edge detector output.  
The dashed line is the TRGB estimate while the dotted line is the 
theoretical fiducial value of --4.0.  Models shown are single-age bursts 
with [Fe/H] = $-1.3$ at 2 Gyr (column 1), 1.78 Gyr (column 2),
1.58 Gyr (column 3), and 1.41 Gyr (column 4).}
\end{figure}

\clearpage
\begin{figure}
\figurenum{7}
\epsscale{1.0}
\plotone{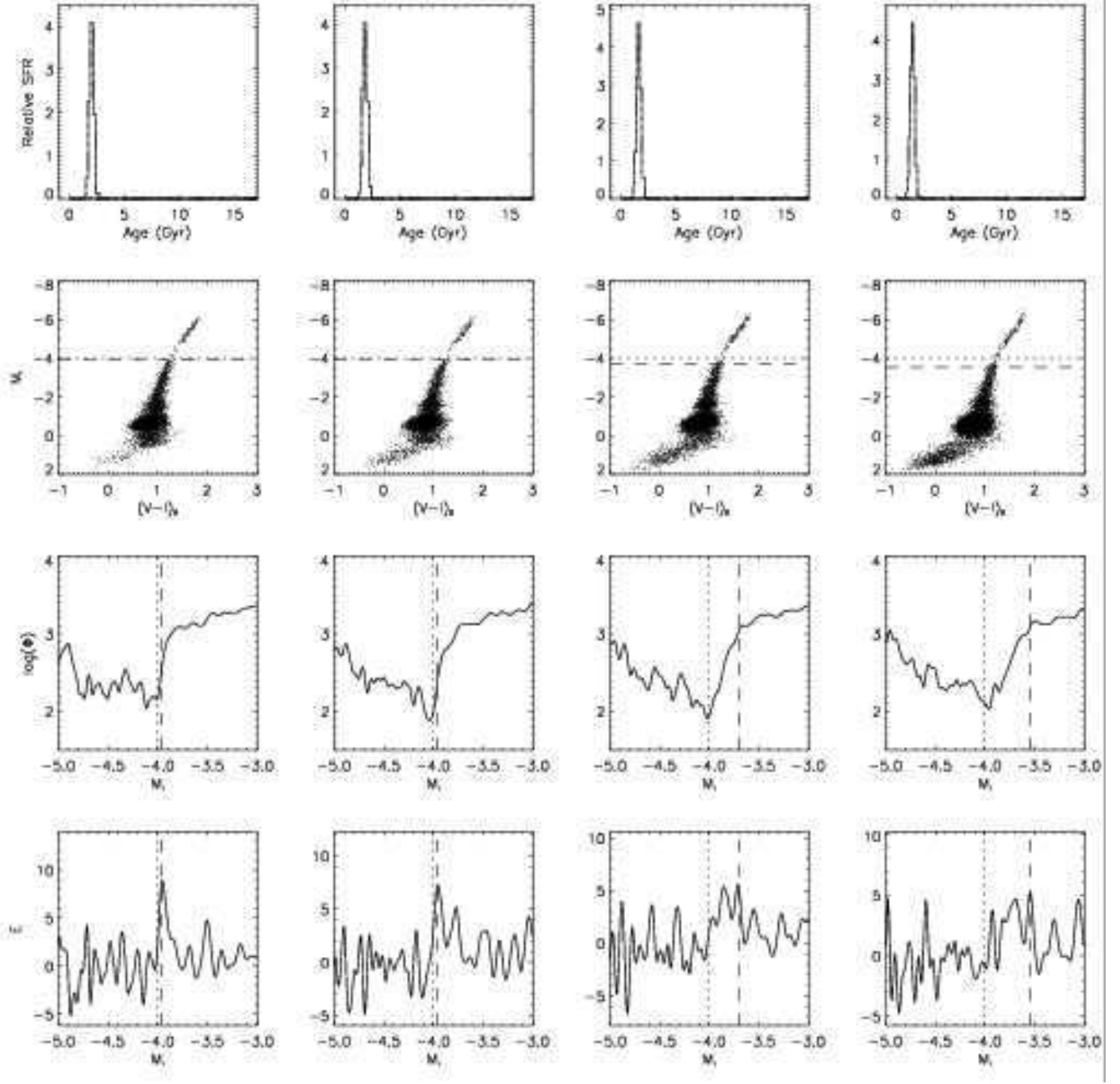}
\caption{Same panels as Fig.\ 6.  Models shown are single 
Gaussian bursts with standard 
deviation of 0.2 Gyr and [Fe/H] = $-1.3$ centered at 2 Gyr 
(column 1), 1.78 Gyr (column 2), 1.58 Gyr 
(column 3), and 1.41 Gyr (column 4).}
\end{figure}

\clearpage
\begin{figure}
\figurenum{8}
\epsscale{1.0}
\plotone{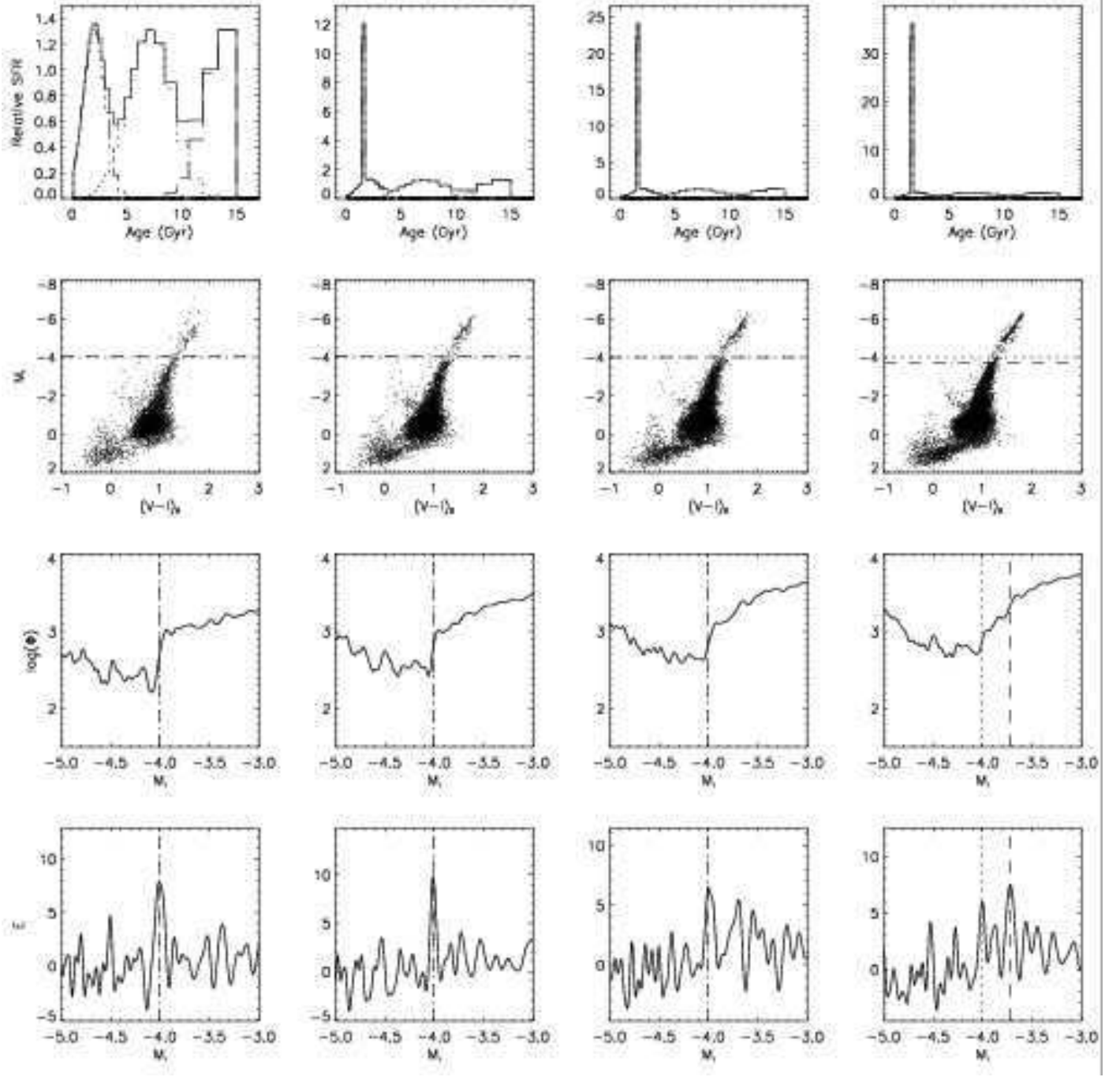}
\caption{Same panels as Fig.\ 6.  Column 1:  Three 
Gaussian bursts at ages 14.12 Gyr, 7.08 Gyr, and
2.00 Gyr with standard deviations of 2, 2, 
and 1 Gyr and metallicities of -1.7, -1.4, and -1.3, respectively.  
The SFR from $1.33-1.68$ Gyr is $\sim 12$ (column 2),
$\sim 25$ (column 3), and $\sim 35$ (column 4) times the
average background rate.}
\end{figure}

\clearpage
\begin{figure}
\figurenum{9}
\epsscale{1.0}
\plotone{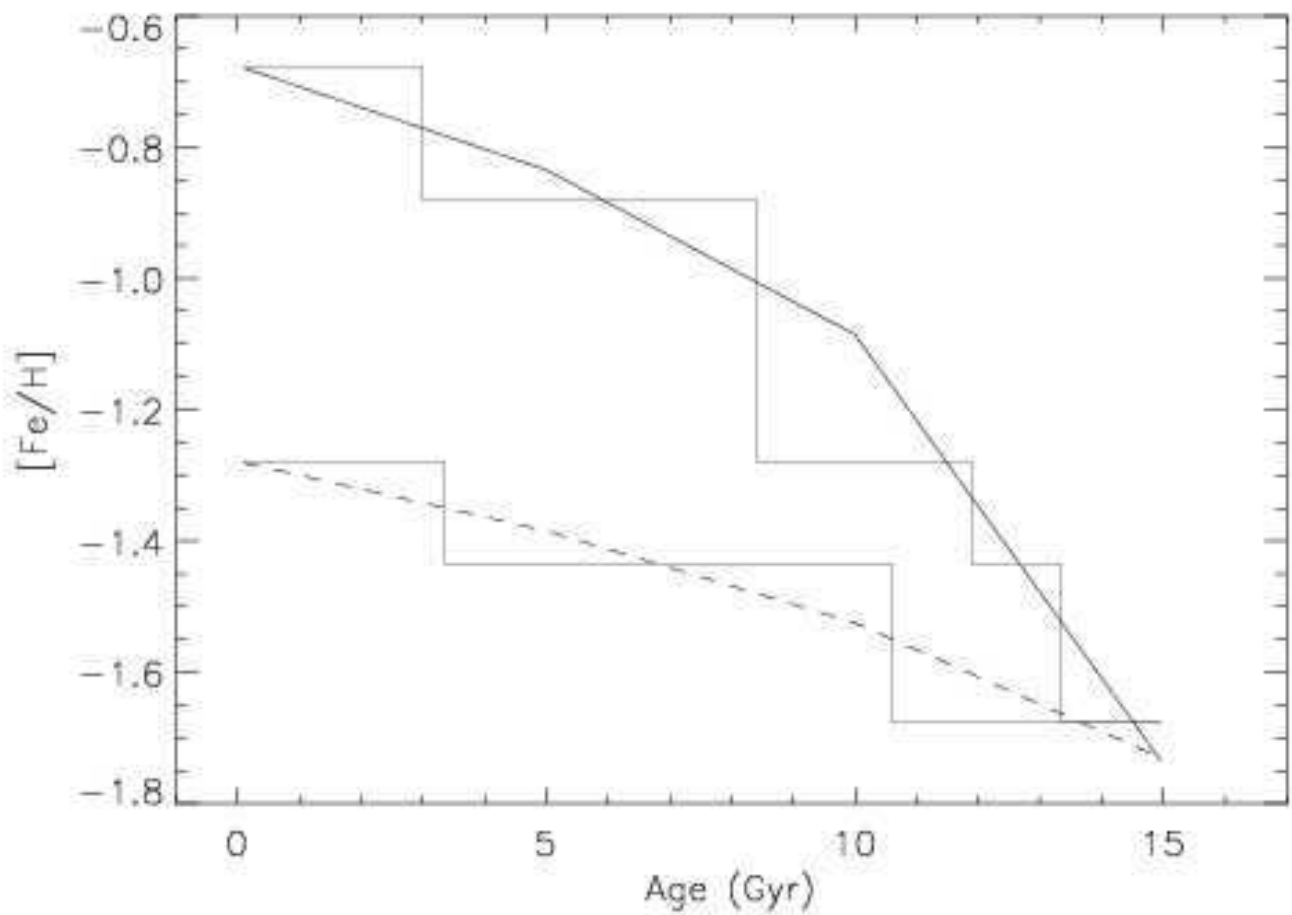}
\caption{Chemical enrichment law \#1 (solid line) and \#2 (dashed line) with
their respective discrete approximations.}
\end{figure}

\clearpage
\begin{figure}
\figurenum{10}
\epsscale{1.0}
\plotone{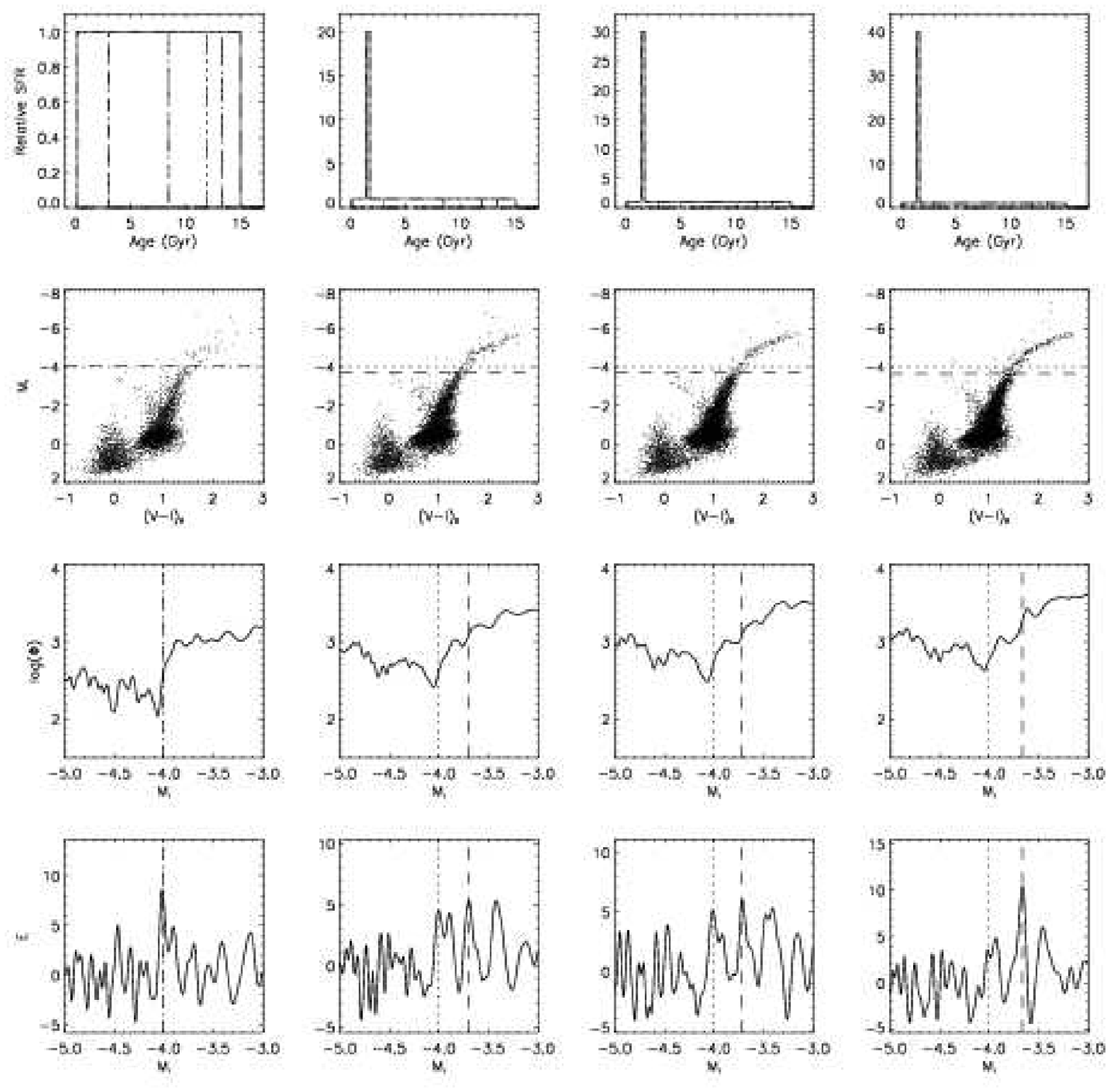}
\caption{Same panels as Fig.\ 6.  Column 1:  Constant SFR following 
CEL \#1.  The dashed vertical 
lines denote the boundaries of different metal abundance.  Starting from the 
oldest, they are [Fe/H] = $-1.7$, $-1.4$, $-1.3$, $-0.9$, $-0.7$.
The young burst from $1.33-1.68$ Gyr is 20 (column 2), 30 (column 3), 
and 40 (column 4) times stronger than the background SFR.}
\end{figure}

\clearpage
\begin{figure}
\figurenum{11}
\epsscale{1.0}
\plotone{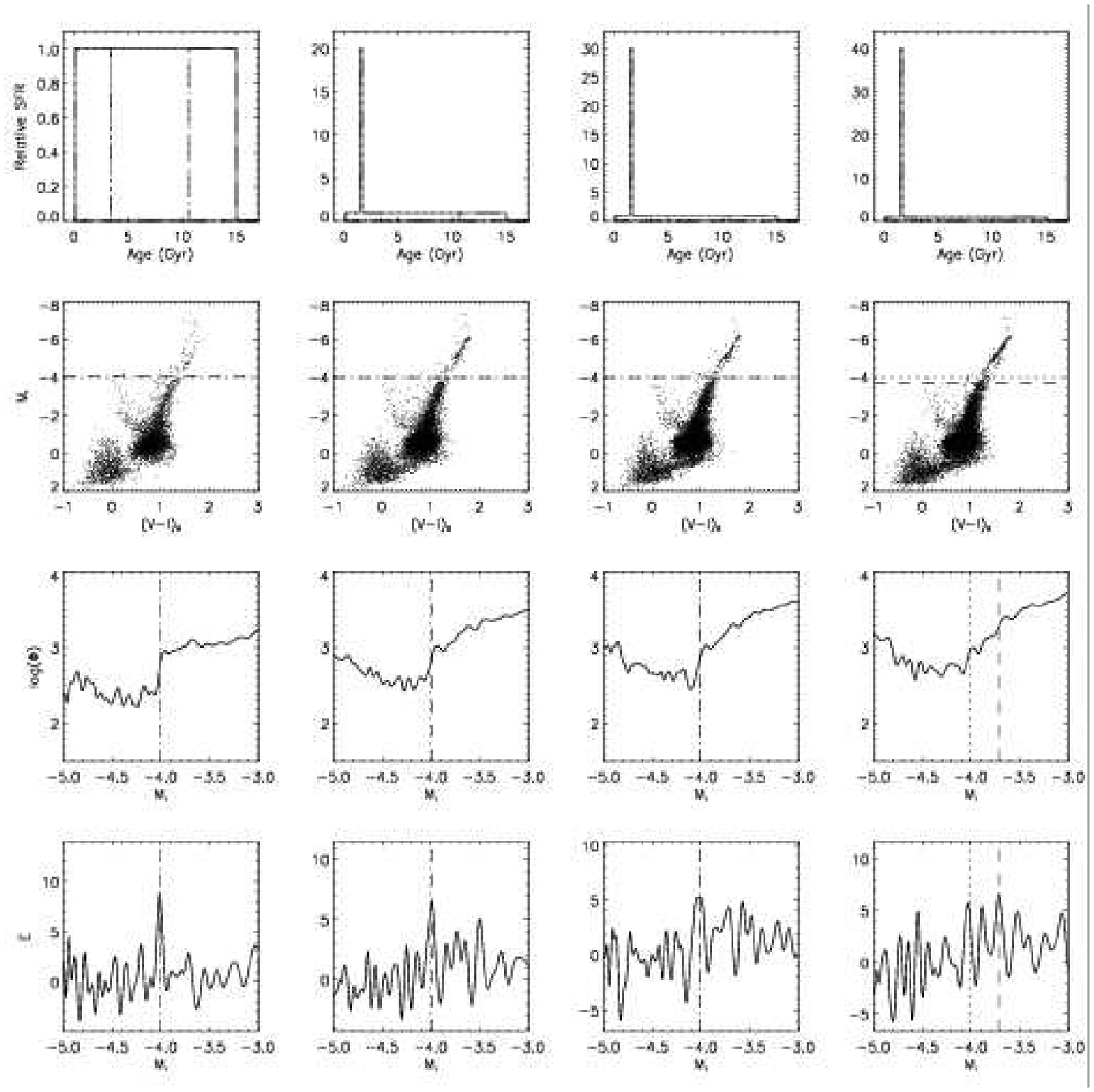}
\caption{Same panels as Fig.\ 6.  Column 1:  Constant SFR following 
CEL \#2.  The dashed vertical lines denote the
boundaries of different metal abundance.  Starting from the oldest, they are
[Fe/H] = $-1.7$, $-1.4$, $-1.3$.  
The young burst from $1.33-1.68$ Gyr is 20 (column 2), 30 (column 3), 
and 40 (column 4) times stronger than the background SFR.}
\end{figure}

\clearpage
\begin{figure}
\figurenum{12}
\epsscale{1.0}
\plotone{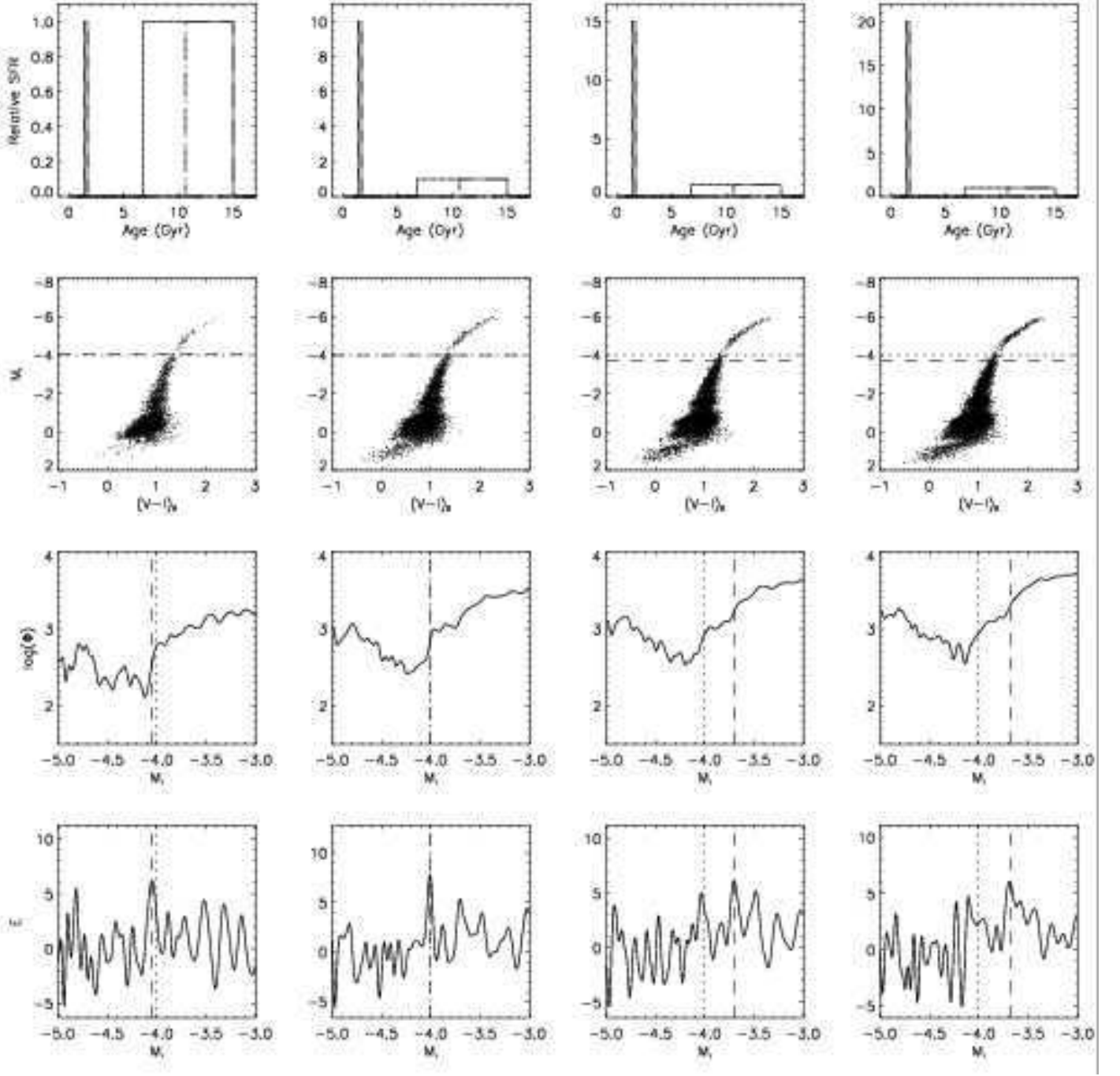}
\caption{Same panels as Fig.\ 6.  Column 1:  Constant SFR.  
The dashed vertical line in top panel denotes the 
boundary of different metal abundance:  [Fe/H] = $-1.7$ 
for ages $>$ 10.59 Gyr, [Fe/H] = $-1.4$ for ages from 
$6.68-10.59$ Gyr and [Fe/H] = $-0.9$ for the 
burst from $1.33-1.68$ Gyr.  
The young burst from $1.33-1.68$ Gyr is 10 (column 2), 15 (column 3), 
and 20 (column 4) times stronger than the background SFR.}
\end{figure}

\clearpage
\begin{figure}
\figurenum{13}
\epsscale{1.0}
\plotone{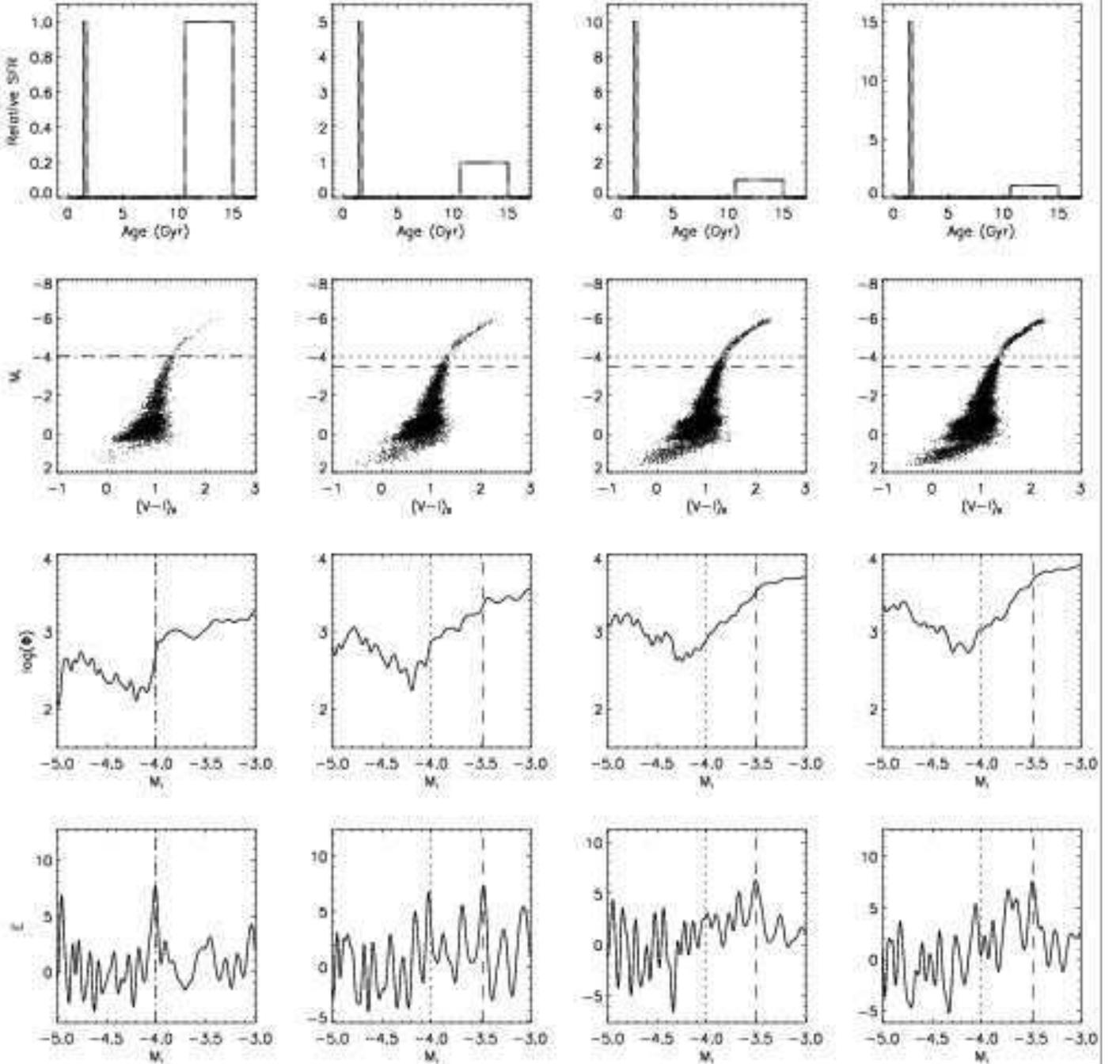}
\caption{Same panels as Fig.\ 6.  Column 1:  Constant SFR.  
The metallicities are [Fe/H] = $-1.7$ for ages $>$ 10.59 Gyr and 
[Fe/H] = $-0.9$ for the 
burst between 1.33 Gyr and 1.68 Gyr.  
The young burst from $1.33-1.68$ Gyr is five (column 2), 10 (column 3), 
and 15 (column 4) times stronger than the background SFR.}
\end{figure}

\clearpage
\begin{figure}
\figurenum{14}
\epsscale{1.0}
\plotone{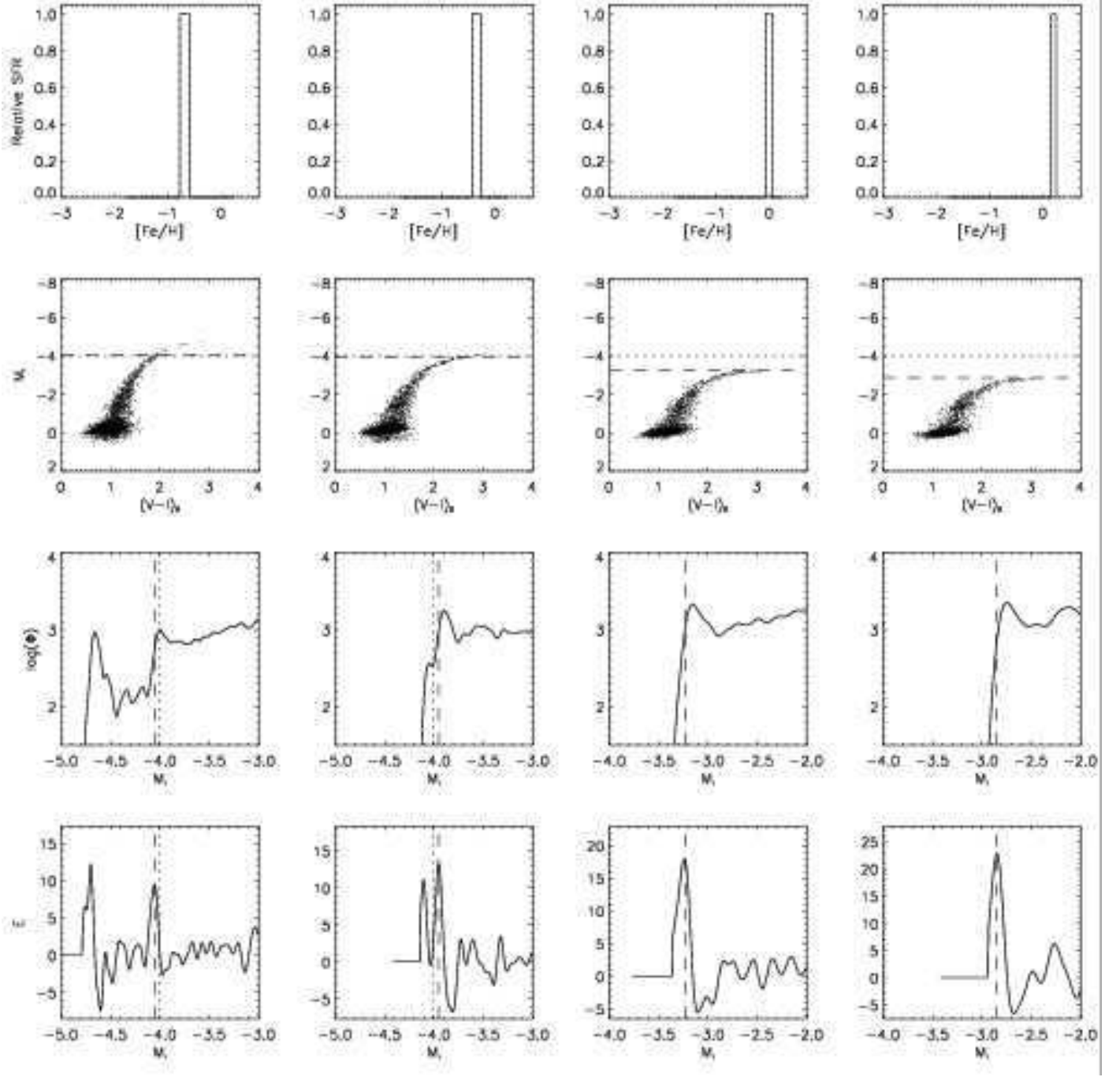}
\caption{Same panels as Fig.\ 6.  Models shown have an age of 
10.00 Gyr and [Fe/H] = -0.7 (column 1), -0.4 (column 2), 
0.0 (column 3), and 0.2 (column 4).}
\end{figure}

\clearpage
\begin{figure}
\figurenum{15}
\epsscale{1.0}
\plotone{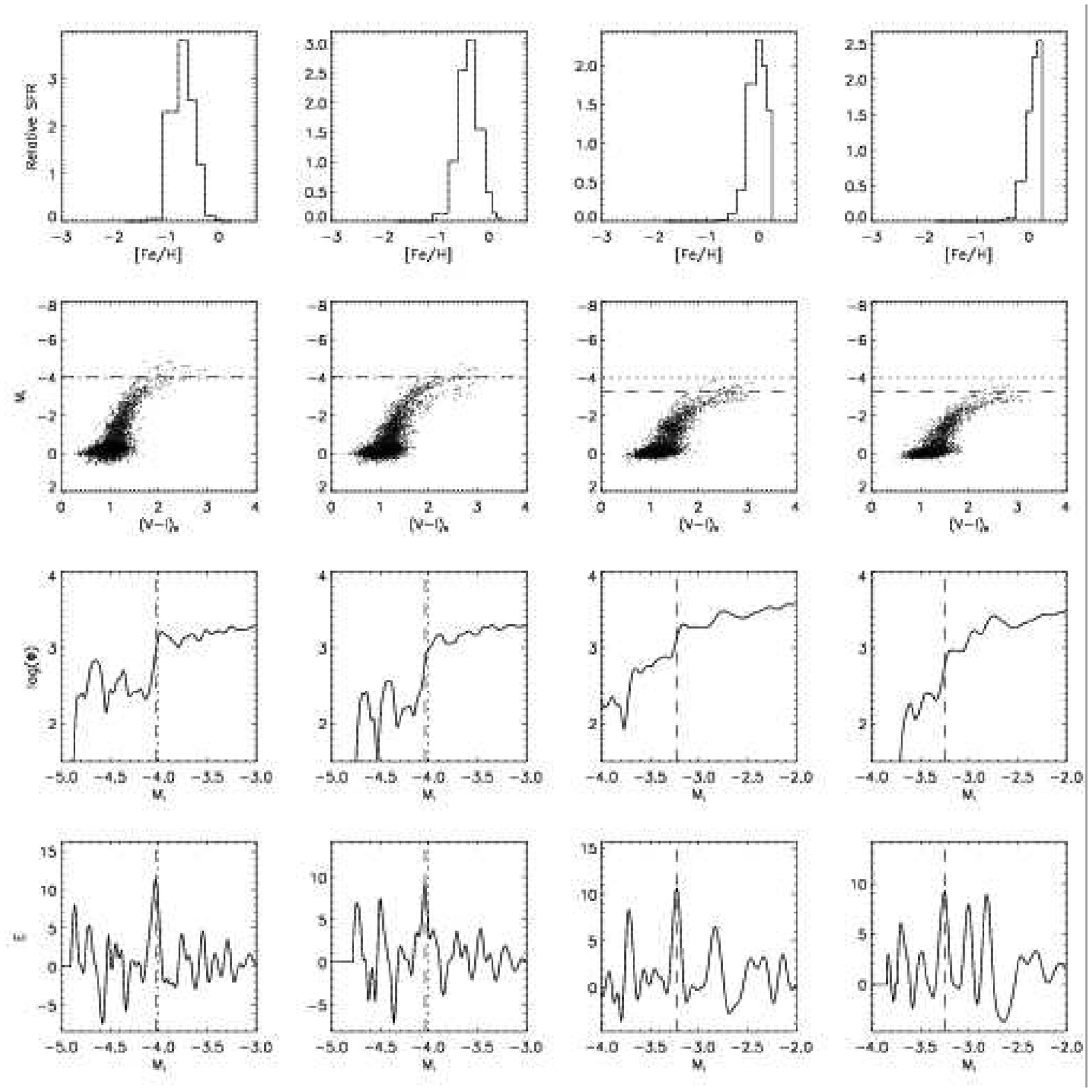}
\caption{Same panels as Fig.\ 6.  Models shown have an age of 
10.00 Gyr and Gaussian metallicity distribution with dispersion of 
0.2 dex centered at [Fe/H] = -0.7 (column 1), -0.4 (column 2), 
0.0 (column 3), 0.2 (column 4).}
\end{figure}

\clearpage
\begin{figure}
\figurenum{16}
\epsscale{1.0}
\plotone{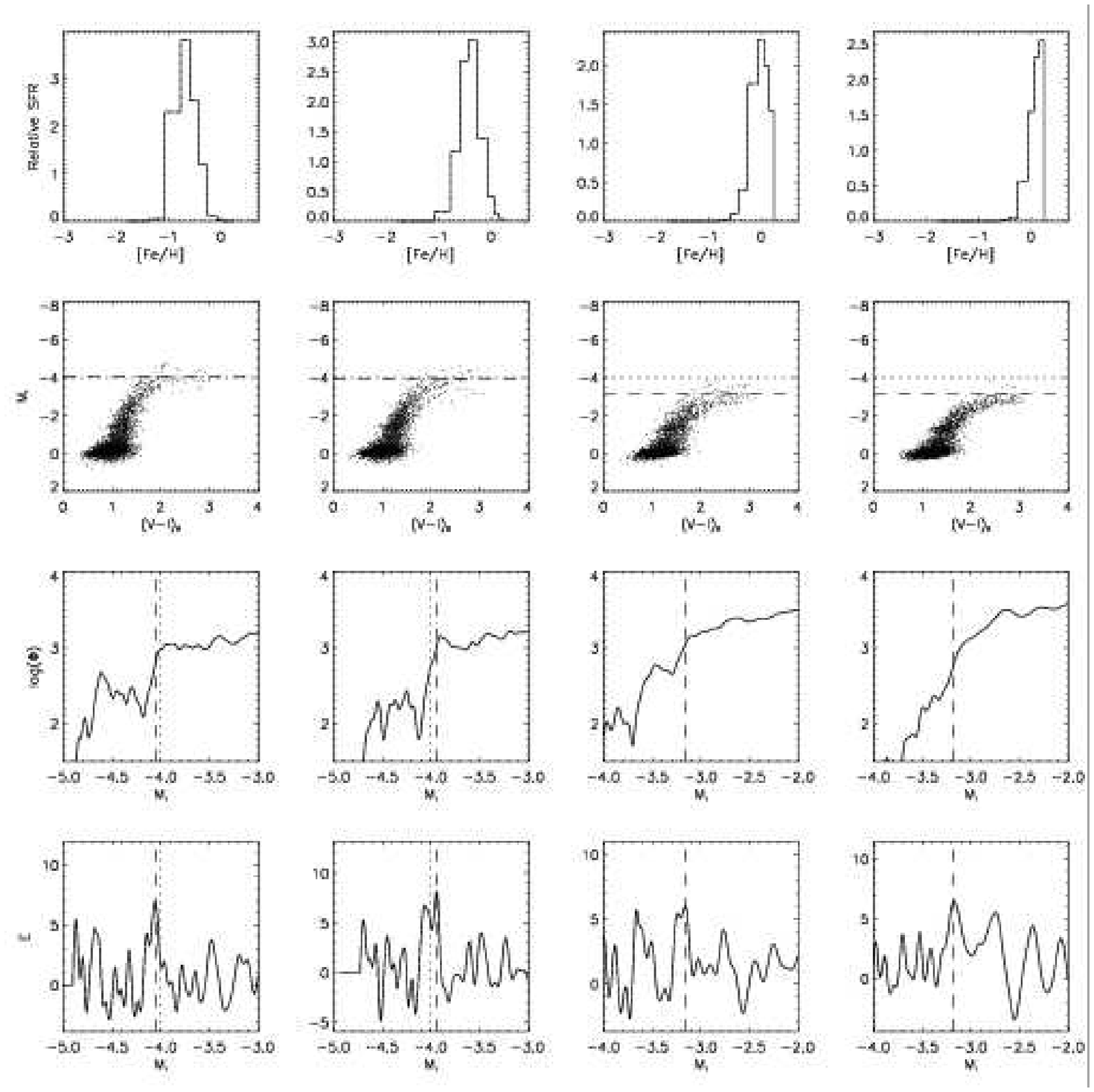}
\caption{Same panels as Fig.\ 6.  Models shown have an age range of 
$9.44-14.96$ Gyr and 
Gaussian metallicity distribution with dispersion of 0.2 dex
centered at [Fe/H] = -0.7 (column 1), -0.4 (column 2), 
0.0 (column 3), 0.2 (column 4).}
\end{figure}

\clearpage
\begin{figure}
\figurenum{17}
\epsscale{1.0}
\plotone{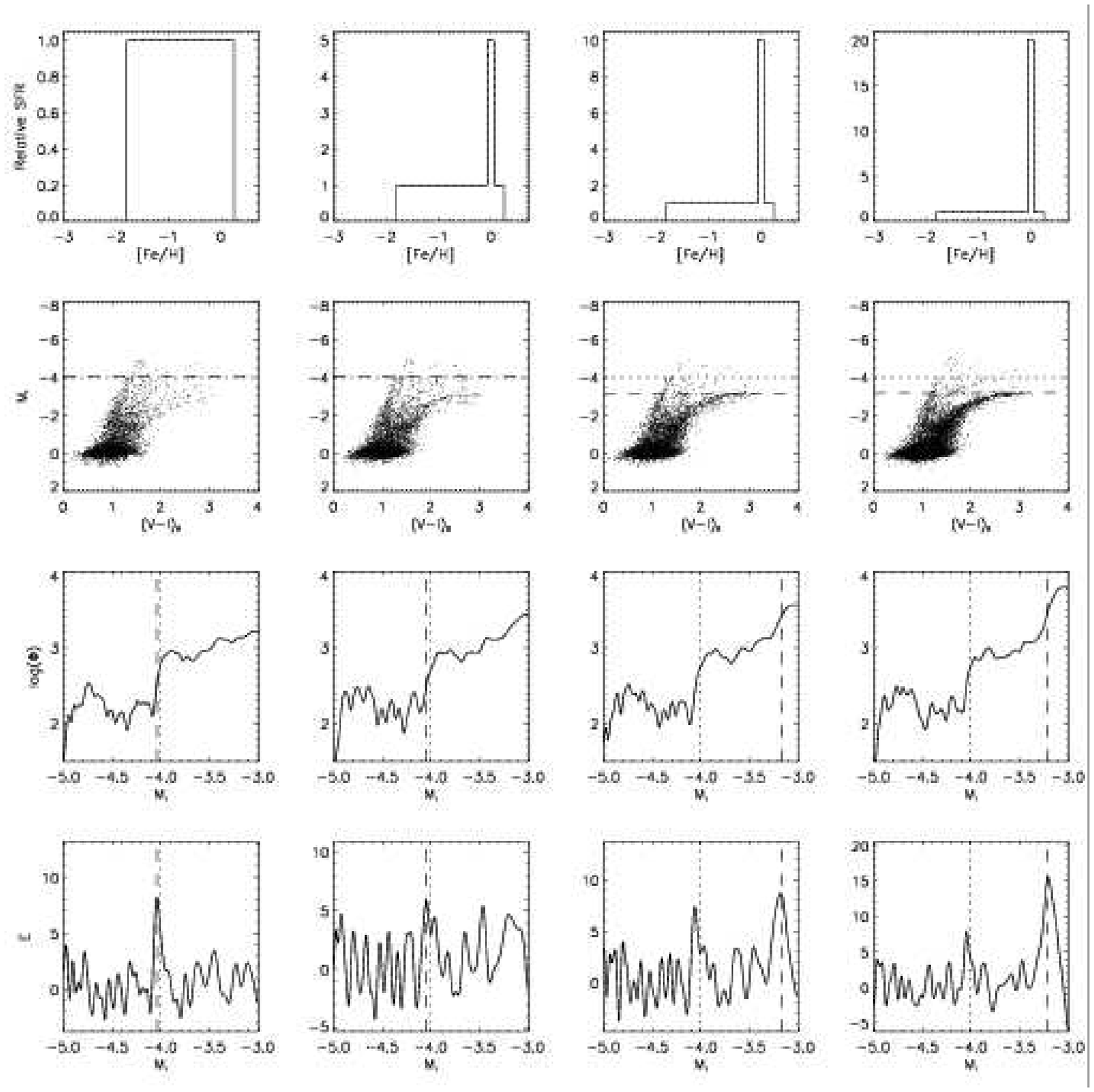}
\caption{Same panels as Fig.\ 6.  Column 1:  Uniform 
metallicity distribution over the range 
$\rm -1.7 \leq [Fe/H] \leq 0.2$ for ages of $9.44-14.96$ Gyr.  
The SFR at [Fe/H] = 0.0 is increased by factors of five (column 2),
10 (column 3), and 20 (column 4).}
\end{figure}

\clearpage
\begin{figure}
\figurenum{18}
\epsscale{1.0}
\plotone{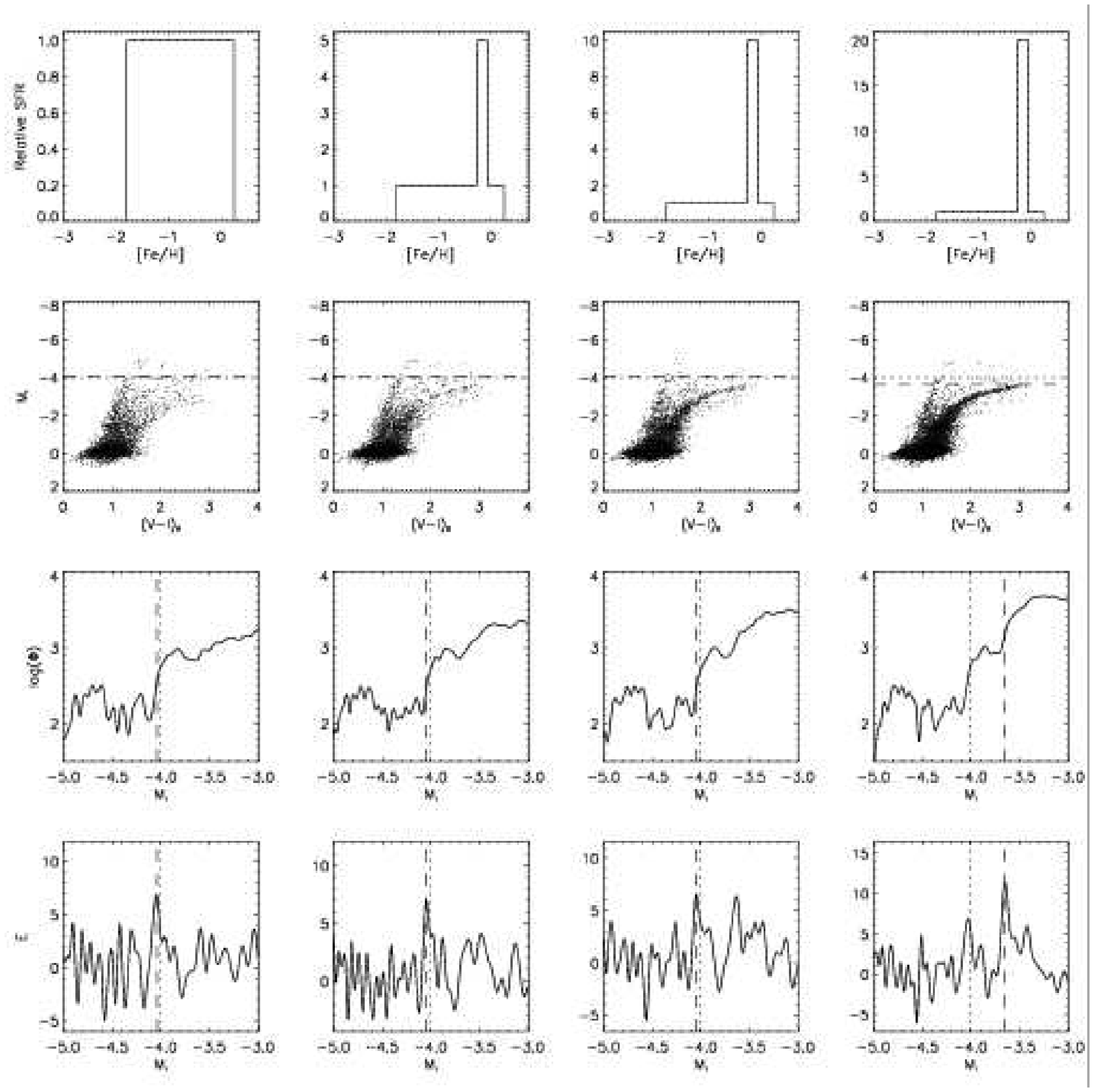}
\caption{Same panels as Fig.\ 6.  Column 1:  Uniform 
metallicity distribution over the range 
$\rm -1.7 \leq [Fe/H] \leq 0.2$ for ages of $9.44-14.96$ Gyr.  
The SFR at [Fe/H] = -0.1 is increased by factors of five (column 2),
10 (column 3), and 20 (column 4).}
\end{figure}

\clearpage
\begin{figure}
\figurenum{19}
\epsscale{1.0}
\plotone{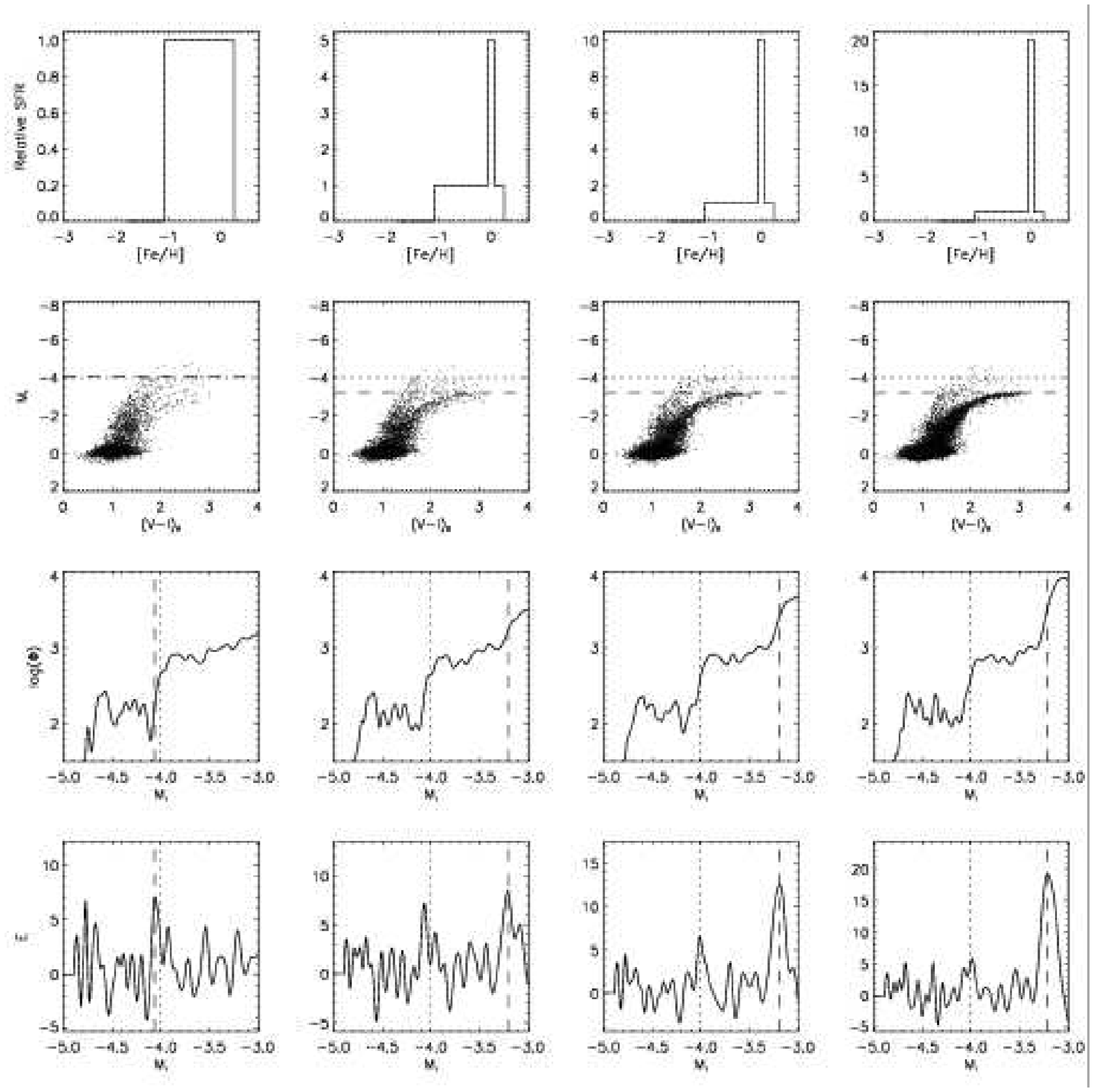}
\caption{Same panels as Fig.\ 6.  Column 1:  Uniform 
metallicity distribution over the range 
$\rm -0.9 \leq [Fe/H] \leq 0.2$ for ages of $9.44-14.96$ Gyr.  
The SFR at [Fe/H] = 0.0 is increased by factors of five (column 2),
10 (column 3), and 20 (column 4).}
\end{figure}

\clearpage
\begin{figure}
\figurenum{20}
\epsscale{1.0}
\plotone{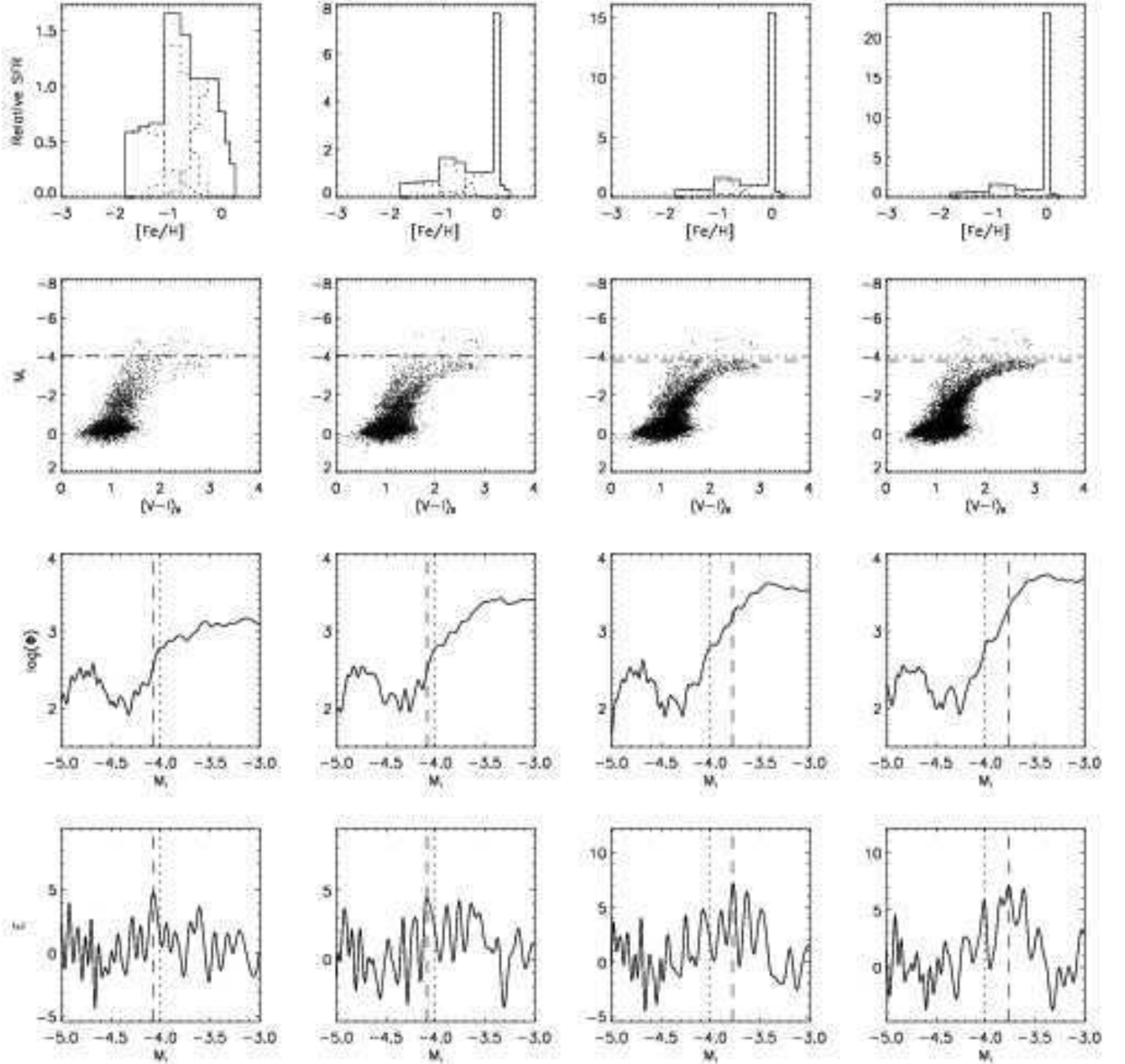}
\caption{Same panels as Fig.\ 6.  Column 1:  The metallicity 
distribution has three
Gaussian components displayed as dotted lines while the total is 
displayed as a solid line.  The metallicities of each component are 
$\rm [Fe/H] = -1.50 \pm 0.45$, $\rm -0.82 \pm 0.20$, and 
$\rm -0.22 \pm 0.26$ while the SFR at each
metallicity is constant over the age ranges $9.44-14.96$, 
$5.96-10.59$, and $2.11-6.68$ Gyr, respectively.
The SFR at [Fe/H] = 0.0 is increased to $\sim 8$ (column 2),
$\sim 15$ (column 3), and $\sim 25$ (column 4) times the average rate.}
\end{figure}

\clearpage
\begin{figure}
\figurenum{21}
\epsscale{1.0}
\plotone{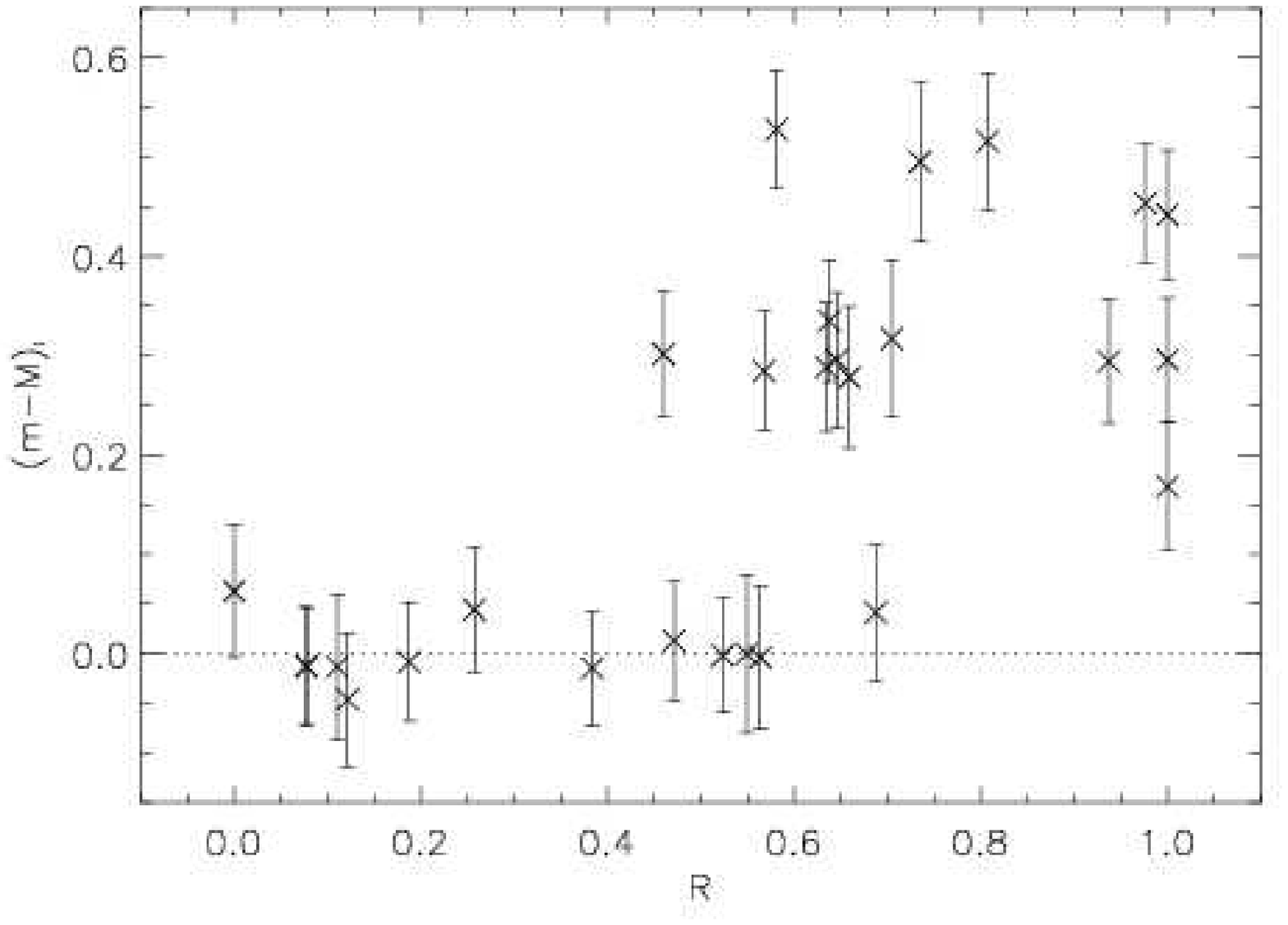}
\caption{Distance modulus assuming $\MITRGB = -4.0$ 
as a function of R which measures the number of young stars near
the TRGB (see text for details).  Error bars denote 
the total error, $\sigma_{tot}$, for each model.  
The dotted line indicates
the distance modulus used for all synthetic CMDs.
For R $>$ 0.60, the dominance of young stars causes errant measurements
of the distance modulus.}
\end{figure}

\clearpage
\begin{figure}
\figurenum{22}
\epsscale{1.0}
\plotone{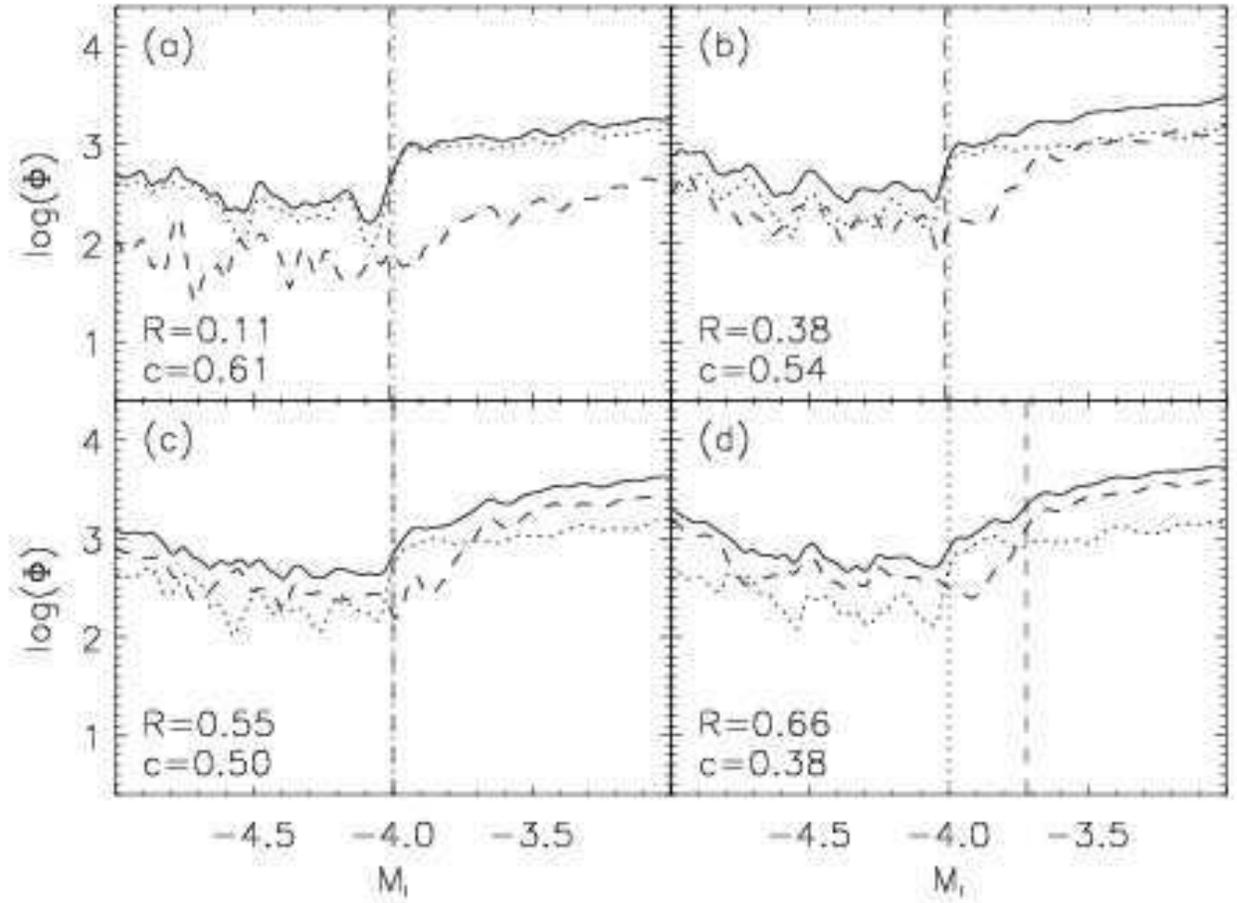}
\caption{Total LF (solid), LF for stars $\geq$ 2 Gyr (dotted) and $<$ 
2 Gyr (dashed) for models discussed in \S 3.3.  Dotted vertical line
corresponds to $\MITRGB = -4.0$ and dashed vertical line is the 
estimated $\MITRGB$.  As R increases from (a) to (d), the LF in the
vicinity of $\MITRGB$ becomes smoother resulting in smaller values of $c$.}
\end{figure}

\clearpage
\begin{figure}
\figurenum{23}
\epsscale{1.0}
\plotone{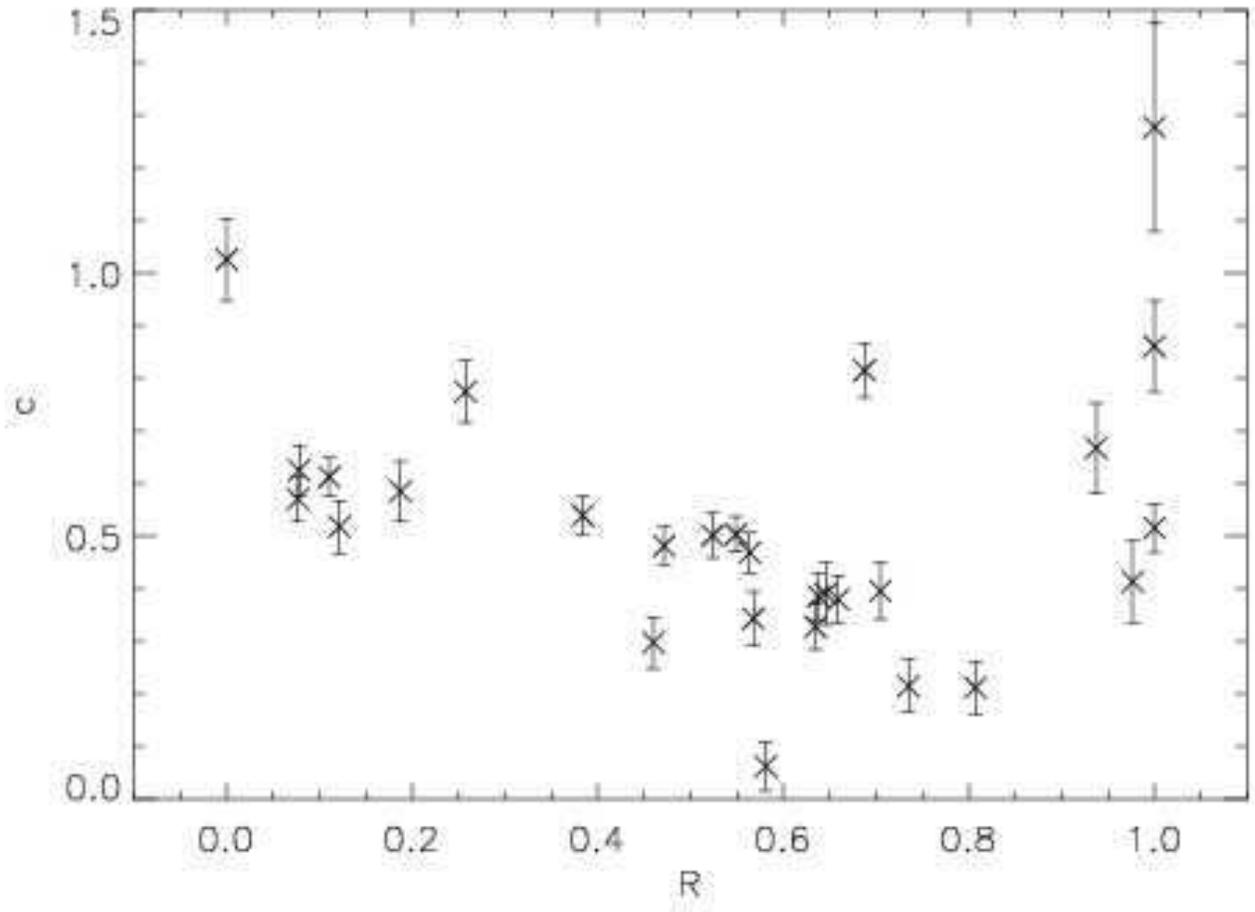}
\caption{Height of the discontinuity, $c$, in the logarithmic LF at the TRGB
as a function of R (see text for details).
Models with a mix of stellar ages (R $<$ 0.9) are negatively 
correlated
at 99.96\% confidence. 
Models with R $>$ 0.9 follow no relationship 
because their LFs near $\MITRGB$ are not smoothed by the 
presence of an older, brighter $\MITRGB$.}
\end{figure}

\clearpage
\begin{figure}
\figurenum{24}
\epsscale{1.0}
\plotone{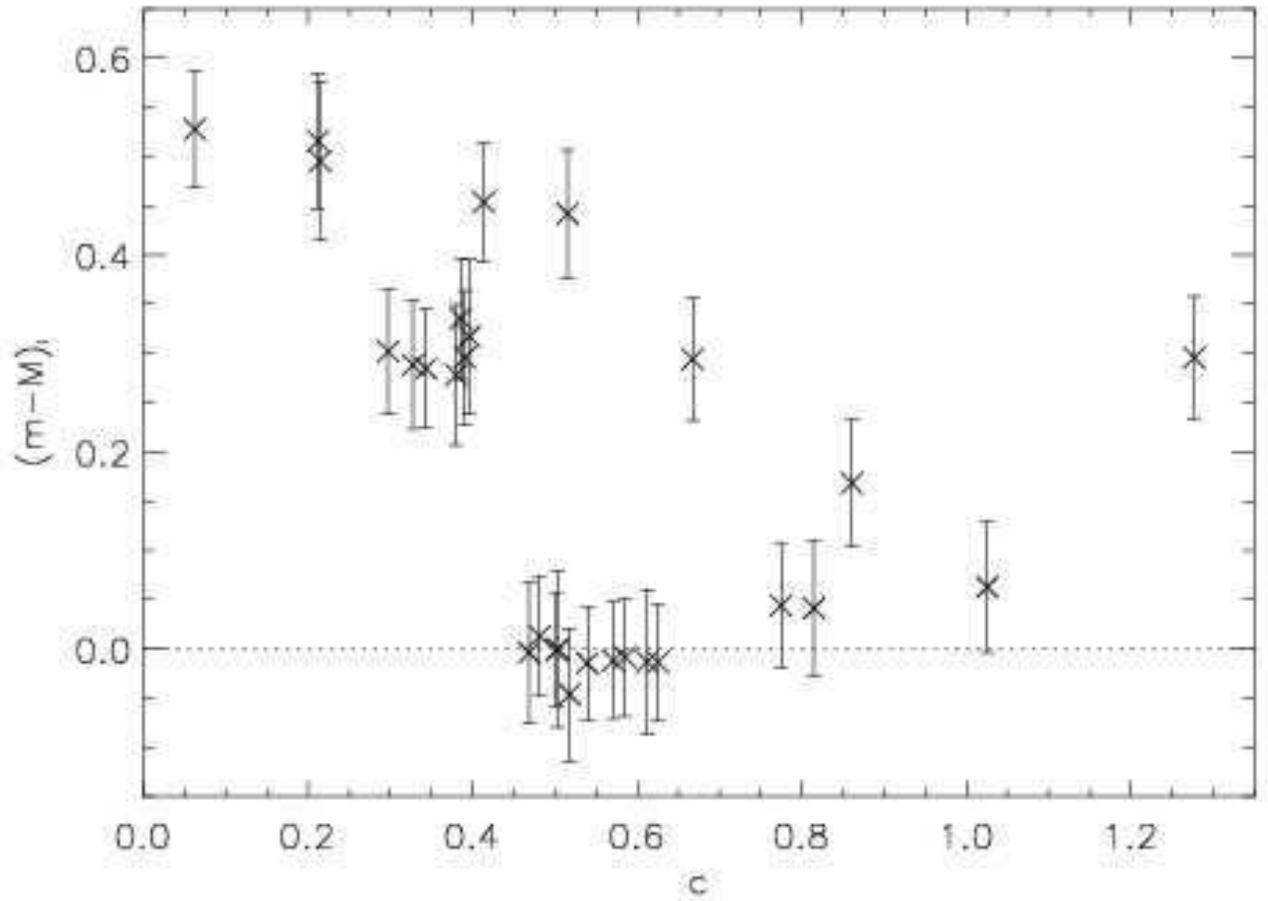}
\caption{Distance modulus as a function of $c$.  
Error bars denote 
the total error, $\sigma_{tot}$, for each model.  
The dotted line indicates
the distance modulus used for all synthetic CMDs.  
Models with a mix of stellar ages give errant measurements of the 
distance 
modulus when $c \lesssim$ 0.40.  Purely young models give errant measurements
regardless of $c$.}
\end{figure}

\clearpage
\begin{figure}
\figurenum{25}
\epsscale{1.0}
\plotone{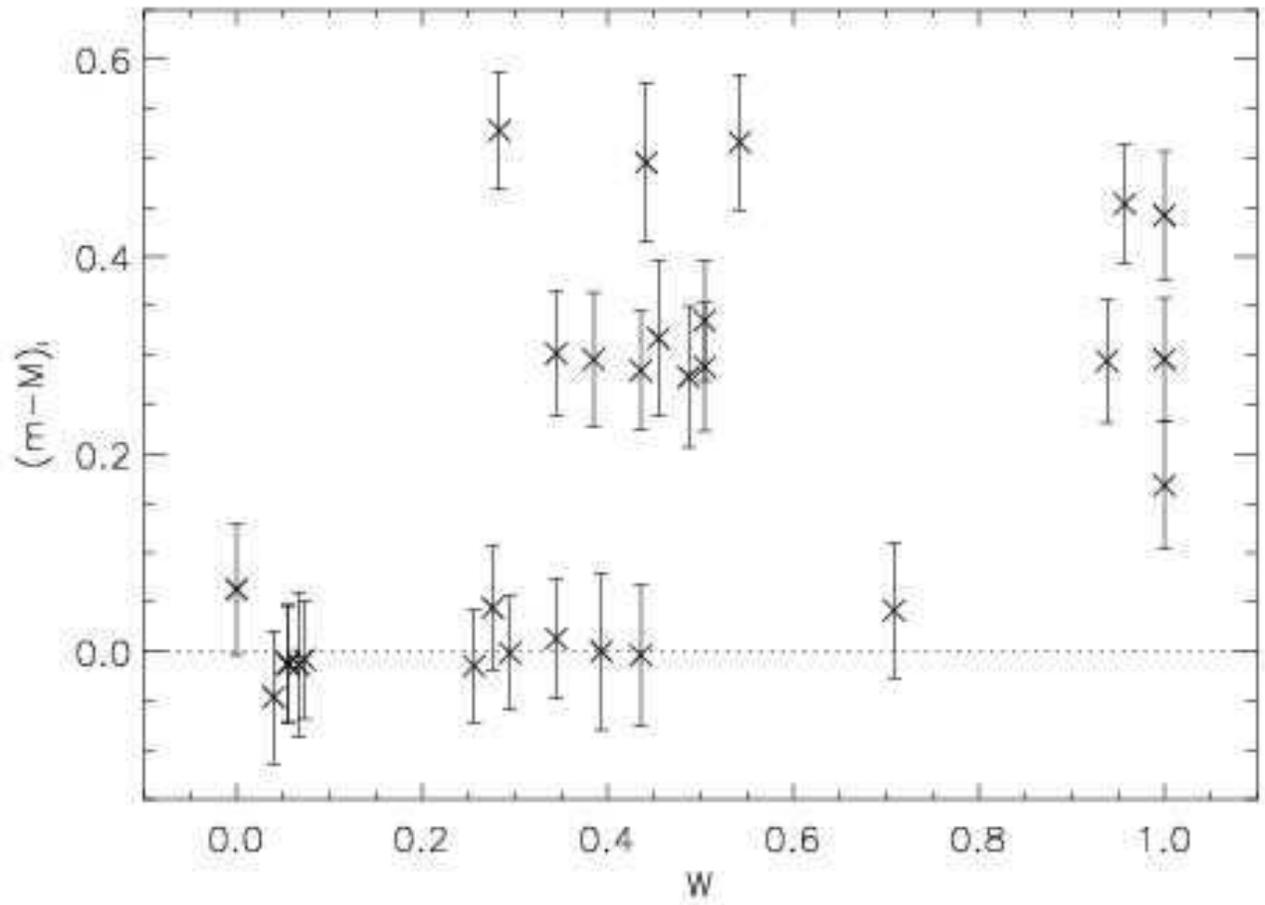}
\caption{Distance modulus as a function of $W$ which measures
the fraction of stars formed over the lifetime of the Universe 
that are formed $1-2$ Gyr ago.
Error bars denote 
the total error, $\sigma_{tot}$, for each model.  
The dotted line indicates
the distance modulus used for all synthetic CMDs.  
For $W$ $>$ 0.30, the number of young stars is large enough to 
cause errant measurements of the distance modulus.}
\end{figure}

\clearpage
\begin{figure}
\figurenum{26}
\epsscale{1.0}
\plotone{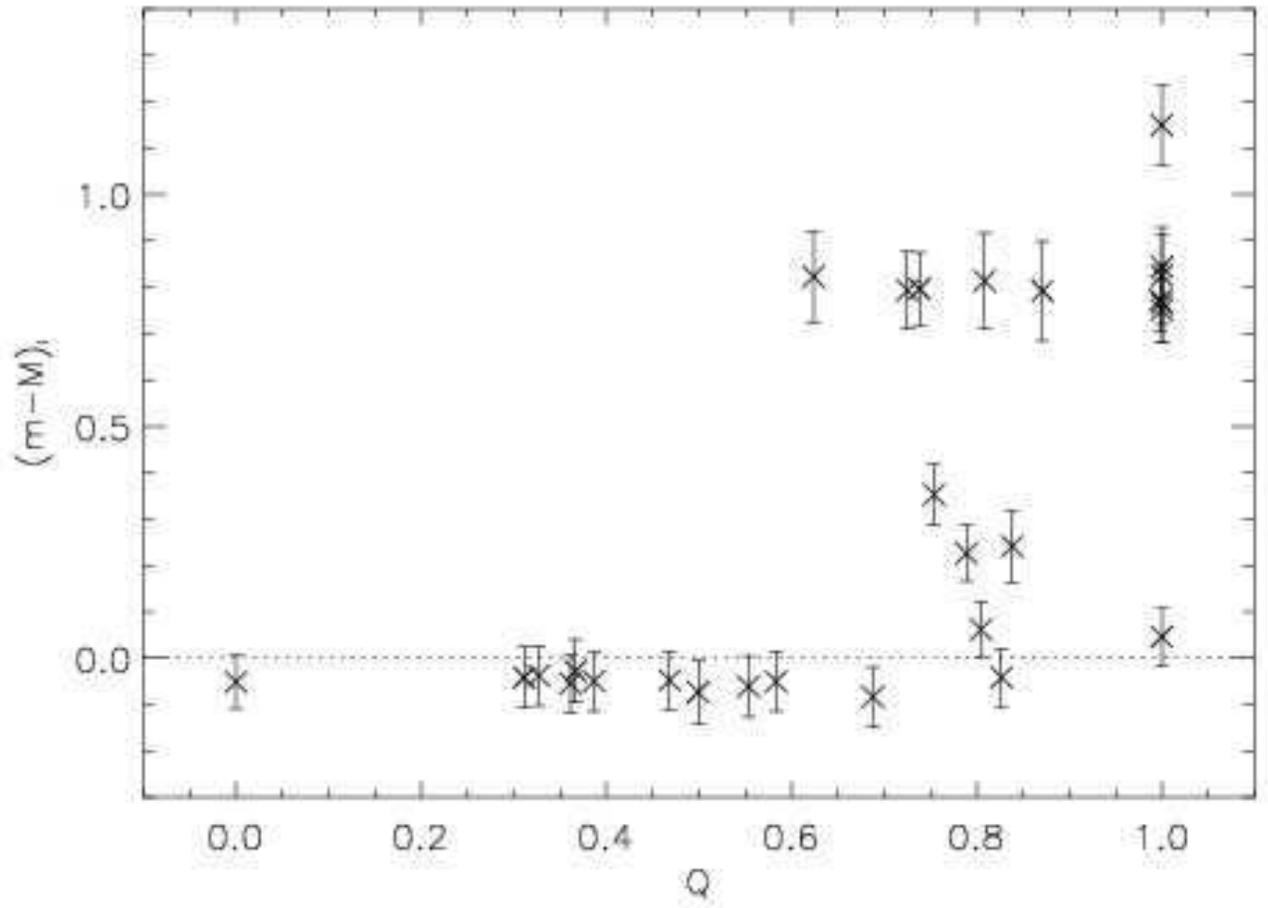}
\caption{Distance modulus as a function of Q which measures
the number of metal-rich stars near the TRGB (see text for details).
Error bars denote 
the total error, $\sigma_{tot}$, for each model.  
The dotted line indicates
the distance modulus used for all synthetic CMDs.
For Q $>$ 0.60, the dominance of metal-rich stars causes errant measurements
of the distance modulus.}
\end{figure}

\clearpage
\begin{figure}
\figurenum{27}
\epsscale{1.0}
\plotone{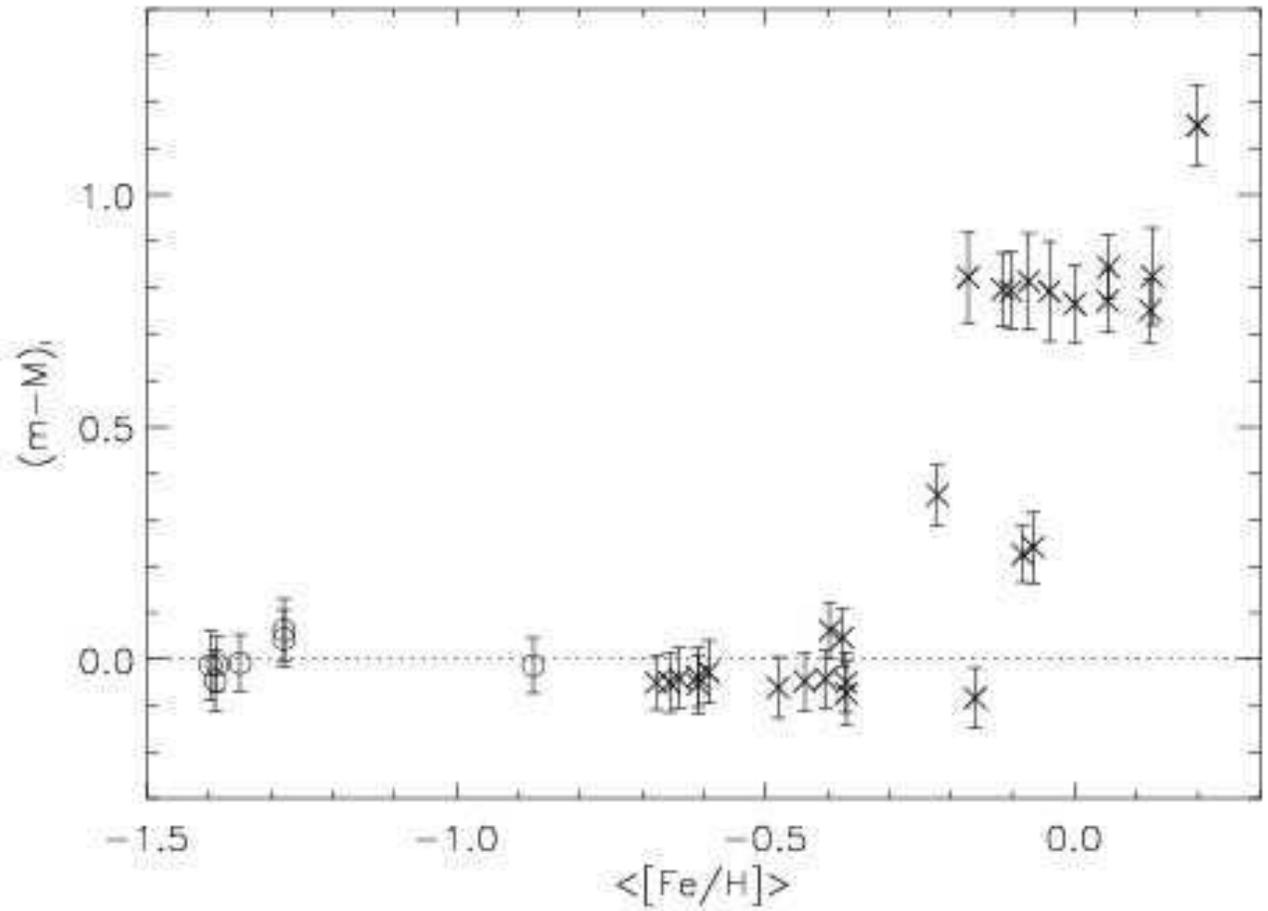}
\caption{Distance modulus as a function of $\rm \langle[Fe/H]\rangle$, 
the average metallicity of stars 0.5 mag below the measured 
TRGB (see text for details).
Error bars denote 
the total error, $\sigma_{tot}$, for each model.  
The dotted line indicates
the distance modulus used for all synthetic CMDs.
Models from column 1 of Figs.\ $6-8$ and $10-13$ are plotted with 
open circles to extend the coverage to lower metallicities.
For $\rm \langle[Fe/H]\rangle > -0.3$, the dominance of metal-rich stars 
causes errant measurements of the distance modulus.}
\end{figure}

\clearpage
\begin{figure}
\figurenum{28}
\epsscale{1.0}
\plotone{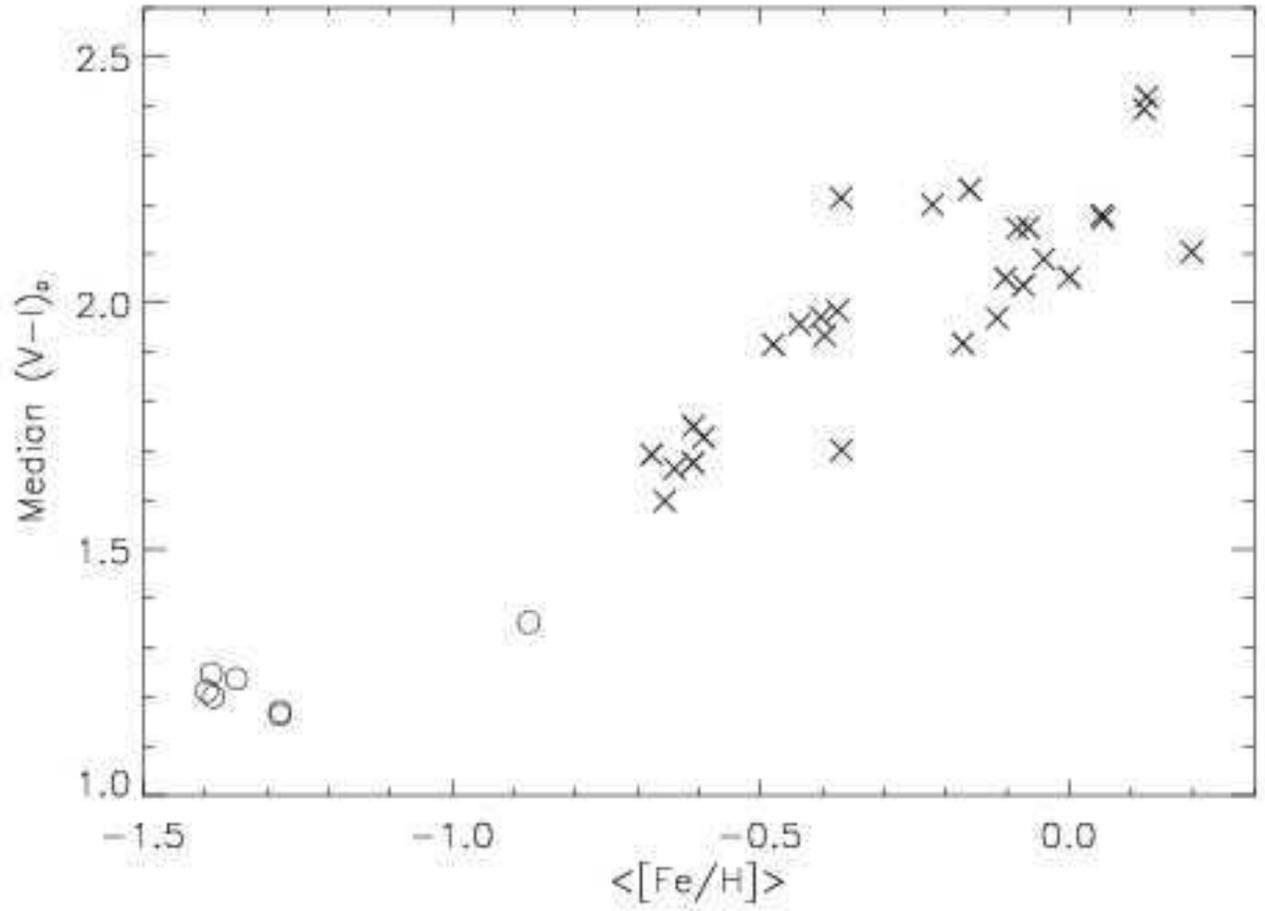}
\caption{Median $\rm (V-I)_0$ as a function of $\rm \langle[Fe/H]\rangle$.  
Models from column 1 of Figs.\ $6-8$ and $10-13$ are plotted with 
open circles to extend the coverage to lower metallicities.  
This confirms the well-known relation between
the color of the TRGB and metallicity.}
\end{figure}

\clearpage
\begin{figure}
\figurenum{29}
\epsscale{1.0}
\plotone{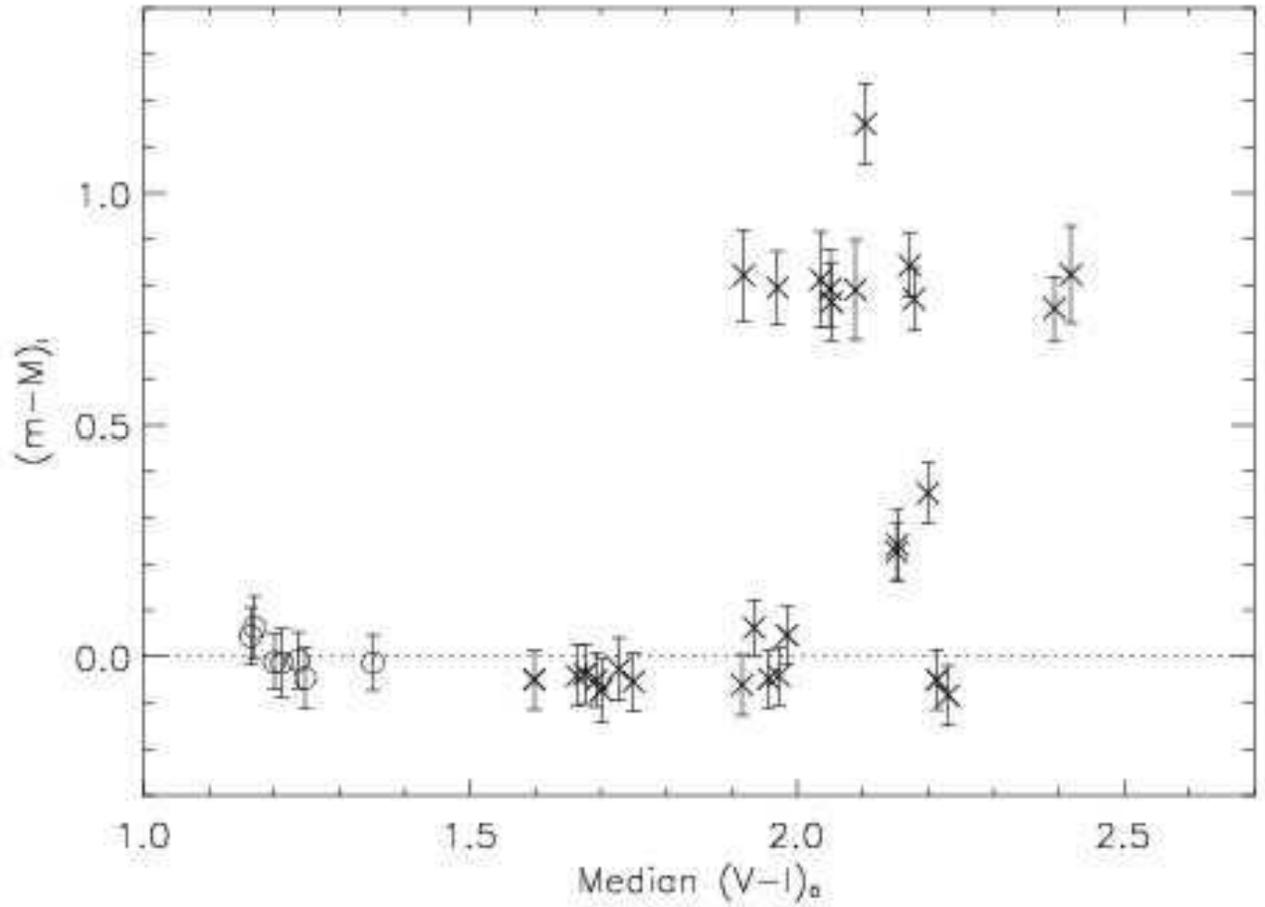}
\caption{Distance modulus as a function of the median $\rm (V-I)_0$ of
stars 0.5 mag below the measured TRGB (see text for details).
Error bars denote 
the total error, $\sigma_{tot}$, for each model.  
The dotted line indicates
the distance modulus used for all synthetic CMDs.
Models from column 1 of Figs.\ $6-8$ and $10-13$ are plotted with 
open circles to extend the coverage to lower metallicities.
For $\rm (V-I)_0 > 1.9$, the dominance of metal-rich stars causes 
errant measurements of the distance modulus.}
\end{figure}

\clearpage
\begin{figure}
\figurenum{30}
\epsscale{1.0}
\plotone{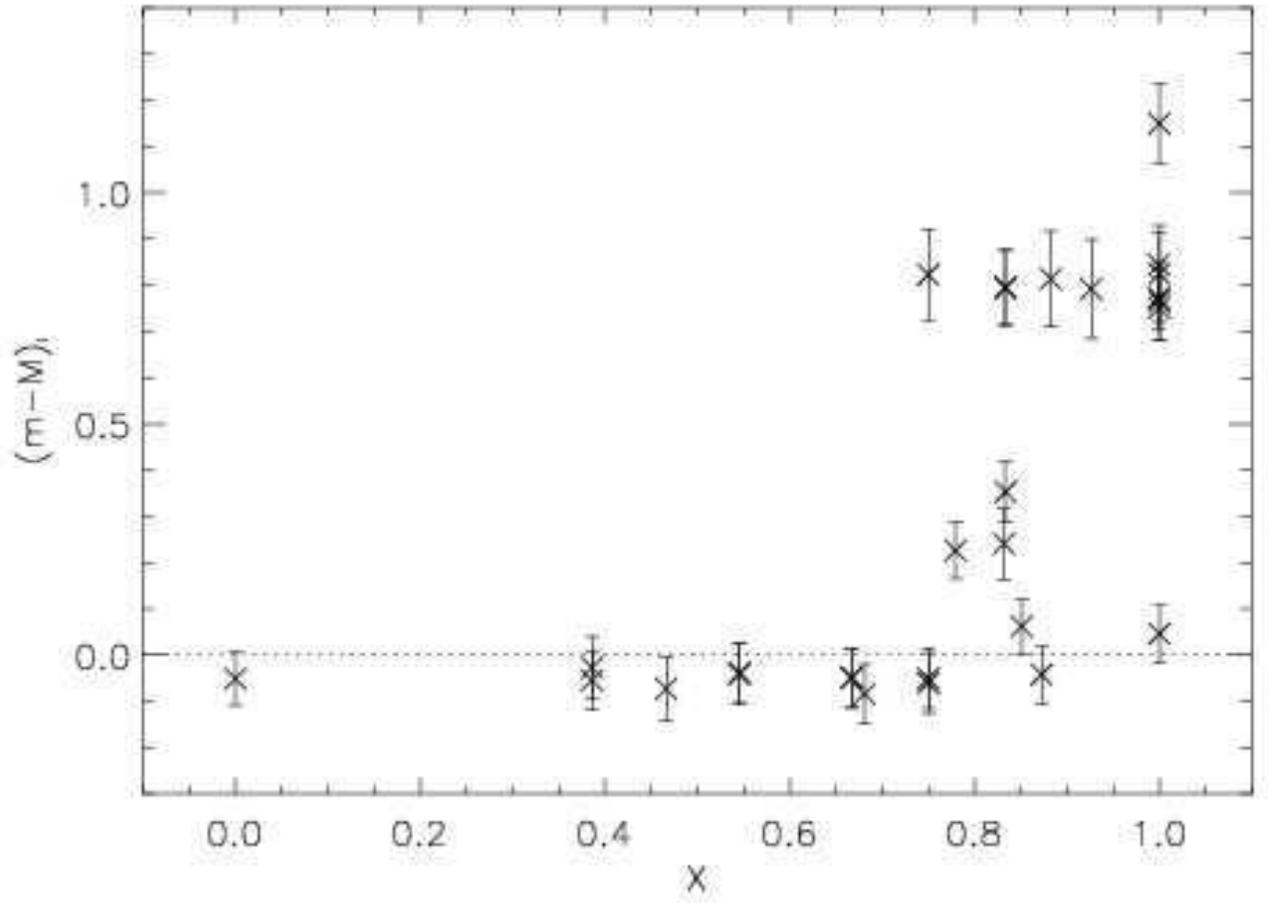}
\caption{Distance modulus as a function of $X$ which measures
the fraction of stars formed at all metallicities 
that are formed with [Fe/H] $> -0.3$.
Error bars denote 
the total error, $\sigma_{tot}$, for each model.  
The dotted line indicates
the distance modulus used for all synthetic CMDs.
For $X >$ 0.70, the dominance of metal-rich stars causes errant measurements
of the distance modulus.}
\end{figure}

\clearpage

\begin{deluxetable}{lccccccrr}
%\begin{deluxetable*}{lccccccrr}
\renewcommand{\arraystretch}{0.7}
\tablenum{1}
\tablecaption{Age Models}
\tablewidth{0pt}
\tablehead{ \colhead{Fig.\#} & \colhead{$\MITRGB$} & 
\colhead{$\sigma_{tot}$} &\colhead{$\sigma_{rand}$} & \colhead{R} 
& \colhead{$c$} & \colhead{$W$}& \colhead{$\rm (m-M)_I$} & \colhead{$\frac{\Delta d}{d}$}\\
\colhead{(Col.\#)} & \colhead{} &\colhead{} & \colhead{} & \colhead{} & \colhead{} &
\colhead{} & \colhead{} & \colhead{}}
\startdata
  6(1)   &-3.94 &0.07 &0.04 &0.00 &1.03 &0.00  & 0.06 & 0.03\\
  6(2)   &-3.83 &0.06 &0.04 &1.00 &0.86 &1.00  & 0.17 & 0.08\\
  6(3)   &-3.70 &0.06 &0.04 &1.00 &1.28 &1.00  & 0.30 & 0.14\\
  6(4)   &-3.56 &0.06 &0.04 &1.00 &0.52 &1.00  & 0.44 & 0.20\\
  7(1)   &-3.96 &0.06 &0.04 &0.26 &0.77 &0.28  & 0.04 & 0.02\\
  7(2)   &-3.96 &0.07 &0.05 &0.69 &0.81 &0.71  & 0.04 & 0.02\\
  7(3)   &-3.71 &0.06 &0.04 &0.94 &0.67 &0.94  & 0.29 & 0.14\\
  7(4)   &-3.55 &0.06 &0.03 &0.98 &0.41 &0.96  & 0.45 & 0.21\\
  8(1)   &-4.01 &0.07 &0.05 &0.11 &0.61 &0.07  &-0.01 &-0.01\\
  8(2)   &-4.01 &0.06 &0.03 &0.38 &0.54 &0.26  &-0.01 &-0.01\\
  8(3)   &-4.00 &0.08 &0.06 &0.55 &0.50 &0.39  & 0.00 & 0.00\\
  8(4)   &-3.72 &0.07 &0.05 &0.66 &0.38 &0.49  & 0.28 & 0.13\\
 10(1)   &-4.01 &0.06 &0.03 &0.08 &0.63 &0.06  &-0.01 &-0.01\\
 10(2)   &-3.70 &0.06 &0.04 &0.46 &0.30 &0.34  & 0.30 & 0.14\\
 10(3)   &-3.72 &0.06 &0.03 &0.57 &0.34 &0.94  & 0.28 & 0.13\\
 10(4)   &-3.66 &0.06 &0.04 &0.64 &0.39 &0.50  & 0.34 & 0.15\\
 11(1)   &-4.01 &0.06 &0.03 &0.08 &0.57 &0.06  &-0.01 &-0.01\\
 11(2)   &-3.99 &0.06 &0.03 &0.47 &0.48 &0.34  & 0.01 & 0.01\\
 11(3)   &-4.00 &0.07 &0.05 &0.56 &0.47 &0.44  & 0.00 &-0.00\\
 11(4)   &-3.71 &0.06 &0.04 &0.63 &0.33 &0.50  & 0.29 & 0.13\\
 12(1)   &-4.05 &0.07 &0.05 &0.12 &0.52 &0.04  &-0.05 &-0.02\\
 12(2)   &-4.00 &0.06 &0.03 &0.52 &0.50 &0.29  & 0.00 & 0.00\\
 12(3)   &-3.70 &0.07 &0.05 &0.65 &0.39 &0.39  & 0.30 & 0.14\\
 12(4)   &-3.68 &0.08 &0.06 &0.70 &0.40 &0.46  & 0.32 & 0.15\\
 13(1)   &-4.01 &0.06 &0.03 &0.19 &0.59 &0.07  &-0.01 & 0.00\\
 13(2)   &-3.47 &0.06 &0.03 &0.58 &0.06 &0.28  & 0.53 & 0.24\\
 13(3)   &-3.50 &0.08 &0.06 &0.73 &0.21 &0.44  & 0.50 & 0.23\\
 13(4)   &-3.48 &0.07 &0.05 &0.81 &0.21 &0.54  & 0.52 & 0.24\\
\enddata
\end{deluxetable}
%\end{deluxetable*}

\clearpage

\begin{deluxetable}{lccccrccrr}
%\begin{deluxetable*}{lccccrccrr}
\renewcommand{\arraystretch}{0.7}
\tablenum{2}
\tablecaption{Metallicity Models}
\tablewidth{0pt}
\tablehead{ \colhead{Fig.\#} & \colhead{$\MITRGB$} & 
\colhead{$\sigma_{tot}$} & \colhead{$\sigma_{rand}$} & \colhead{Q} & 
\colhead{$\rm \langle[Fe/H]\rangle$} &
\colhead{Median} & \colhead{$X$} & \colhead{$\rm (m-M)_I$} & \colhead{$\frac{\Delta d}{d}$}\\
\colhead{(Col.\#)} & \colhead{} & \colhead{} & \colhead{} & \colhead{} & 
\colhead{} & \colhead{$\rm (V-I)_0$} & \colhead{} & \colhead{} & \colhead{}}
\startdata
 14(1)   &-4.05 &0.06 &0.03 &0.00 &-0.68 &1.69 &0.00  &-0.05 &-0.02\\
 14(2)   &-3.95 &0.06 &0.04 &1.00 &-0.38 &1.98 &1.00  & 0.05 & 0.02\\
 14(3)   &-3.24 &0.08 &0.06 &1.00 & 0.00 &2.05 &1.00  & 0.76 & 0.35\\
 14(4)   &-2.85 &0.09 &0.07 &1.00 & 0.20 &2.10 &1.00  & 1.15 & 0.53\\
 15(1)   &-4.03 &0.07 &0.05 &0.37 &-0.59 &1.73 &0.39  &-0.03 &-0.01\\
 15(2)   &-4.04 &0.06 &0.04 &0.83 &-0.40 &1.97 &0.87  &-0.04 &-0.02\\
 15(3)   &-3.23 &0.07 &0.05 &1.00 & 0.05 &2.18 &1.00  & 0.77 & 0.36\\
 15(4)   &-3.25 &0.07 &0.05 &1.00 & 0.12 &2.39 &1.00  & 0.75 & 0.35\\
 16(1)   &-4.05 &0.06 &0.04 &0.36 &-0.61 &1.75 &0.39  &-0.05 &-0.02\\
 16(2)   &-3.94 &0.06 &0.03 &0.80 &-0.40 &1.93 &0.85  & 0.06 & 0.03\\
 16(3)   &-3.16 &0.07 &0.05 &1.00 & 0.05 &2.17 &1.00  & 0.84 & 0.39\\
 16(4)   &-3.18 &0.10 &0.09 &1.00 & 0.13 &2.42 &1.00  & 0.82 & 0.38\\
 17(1)   &-4.04 &0.06 &0.04 &0.33 &-0.61 &1.68 &0.54  &-0.04 &-0.02\\
 17(2)   &-4.05 &0.06 &0.04 &0.39 &-0.66 &1.60 &0.67  &-0.05 &-0.02\\
 17(3)   &-3.18 &0.10 &0.08 &0.62 &-0.17 &1.92 &0.75  & 0.82 & 0.38\\
 17(4)   &-3.21 &0.08 &0.07 &0.73 &-0.10 &2.05 &0.83  & 0.79 & 0.37\\
 18(1)   &-4.04 &0.07 &0.04 &0.31 &-0.64 &1.66 &0.54  &-0.04 &-0.02\\
 18(2)   &-4.05 &0.06 &0.04 &0.47 &-0.44 &1.96 &0.67  &-0.05 &-0.02\\
 18(3)   &-4.05 &0.06 &0.04 &0.58 &-0.37 &2.21 &0.75  &-0.05 &-0.02\\
 18(4)   &-3.65 &0.06 &0.04 &0.75 &-0.22 &2.20 &0.83  & 0.35 & 0.16\\
 19(1)   &-4.06 &0.07 &0.04 &0.55 &-0.48 &1.92 &0.75  &-0.06 &-0.03\\
 19(2)   &-3.20 &0.08 &0.06 &0.74 &-0.12 &1.97 &0.83  & 0.80 & 0.37\\
 19(3)   &-3.19 &0.10 &0.09 &0.81 &-0.07 &2.04 &0.88  & 0.81 & 0.37\\
 19(4)   &-3.21 &0.11 &0.09 &0.87 &-0.04 &2.09 &0.92  & 0.79 & 0.36\\
 20(1)   &-4.07 &0.07 &0.05 &0.50 &-0.37 &1.70 &0.47  &-0.07 &-0.03\\
 20(2)   &-4.08 &0.07 &0.04 &0.69 &-0.16 &2.23 &0.68  &-0.08 &-0.04\\
 20(3)   &-3.77 &0.06 &0.03 &0.79 &-0.08 &2.15 &0.78  & 0.23 & 0.10\\
 20(4)   &-3.76 &0.08 &0.06 &0.84 &-0.07 &2.15 &0.83  & 0.24 & 0.11\\
  6(1)   &-3.94 &0.07 &0.04 &0.00 &-1.28 &1.17 &0.00  & 0.06 & 0.03\\
  7(1)   &-3.96 &0.06 &0.04 &0.00 &-1.28 &1.16 &0.00  & 0.04 & 0.02\\
  8(1)   &-4.01 &0.07 &0.05 &0.00 &-1.40 &1.21 &0.00  &-0.01 &-0.01\\
 10(1)   &-4.01 &0.06 &0.03 &0.00 &-0.88 &1.35 &0.00  &-0.01 &-0.01\\
 11(1)   &-4.01 &0.06 &0.03 &0.00 &-1.39 &1.20 &0.00  &-0.01 &-0.01\\
 12(1)   &-4.05 &0.07 &0.05 &0.00 &-1.39 &1.25 &0.00  &-0.05 &-0.02\\
 13(1)   &-4.01 &0.06 &0.03 &0.00 &-1.35 &1.24 &0.00  &-0.01 & 0.00\\
\enddata
\tablecomments{The last
seven models are included from Table 1 to extend the coverage to lower
metallicities.}
\end{deluxetable}
%\end{deluxetable*}

\end{document}